\documentclass[a4paper,11pt]{article}
\usepackage[a4paper,
            bindingoffset=0.2in,
            left=1in,
            right=1in,
            top=1in,
            bottom=1in,
            footskip=.5in]{geometry}
\usepackage[T1]{fontenc}
\usepackage{inputenc}
\usepackage{amssymb} 
\usepackage{amsfonts}
\usepackage[labelsep=endash]{caption}
\usepackage{subcaption}
\usepackage{arydshln}
\usepackage{mathtools}

\newcommand{\footremember}[2]{%
\footnote{#2}
\newcounter{#1}
\setcounter{#1}{\value{footnote}}%
}
\newcommand{\footrecall}[1]{%
\footnotemark[\value{#1}]%
}
\newcommand*{\affaddr}[1]{#1} % No op here. Customize it for different styles.

\usepackage{amsmath,bm}
\usepackage{mathptmx}
\usepackage{booktabs}
\usepackage{times}
\providecommand{\keywords}[1]{\textbf{\textit{Keywords:}} #1}
\usepackage{hyperref}
\usepackage{multirow}
\usepackage{multicol} % used for the two-column index
\usepackage{longtable}
\usepackage{graphicx}
\usepackage{floatrow}
\usepackage{csquotes}
\usepackage[square,numbers]{natbib}
\title{Small Area Estimation of Household Economic Indicators under Unit-Level Generalized Additive Models for Location, Scale and Shape}
\date{}
\author{%
  Lorenzo Mori\footremember{alley} {\affaddr{University of Bologna. Dep. of Statistical Sciences P. Fortunati. Bologna, Italy. Corresponding author Email and ORCID: \href{mailto:maria.ferrante@unibo.it}{maria.ferrante@unibo.it}} \url{https://orcid.org/0000-0001-9813-2420}}%
  \and  Maria Rosaria Ferrante\footrecall{alley} %
  }
\begin{document}
\maketitle
\begin{abstract}
We propose a Small Area Estimation model based on Generalized Additive Models for Location, Scale and Shape (SAE-GAMLSS), for the estimation of household economic indicators. SAE-GAMLSS release the exponential family distributional assumption and allow each distributional parameter to depend on covariates. A bootstrap approach to estimate MSE is proposed. The SAE-GAMLSS estimator shows a largely better performance than the well-known EBLUP, under various simulated scenarios. Based on SAE-GAMLSS per-capita consumption of Italian and foreign households in Italian regions, in urban and rural areas, is estimated. Results show that the well-known Italian North-South divide does not hold for foreigners.
\end{abstract}
\keywords{GAMLSS, kurtosis, per-capita expenditure, skewness, urban-rural disparity}

\section{Introduction}
\label{sec1}
Sample surveys are designed to provide reliable estimates for the whole reference population and, occasionally, for large sub-populations. For specific sub-populations, the sample size could be too small to produce design-based estimates with an acceptable level of variability. In this case, small area estimation (SAE) models are used to estimate sub-population parameters, where sub-population can be defined using geographical areas and/or socio-demographic categories. The “small” term then, linked to “area” or to “domain”, does not refer to the size of the population, but to the domain sample size. SAE models try to reduce estimator variability for a certain area by ‘borrowing strength’ from neighbouring areas and from auxiliary information available from administrative or census data. They can be roughly classified in two groups, area-level and unit-level models (for an overview of SAE methods see \citep{rao_small_2015}). Area-level models are used when only area-level auxiliary information, generally less subject to confidentiality restrictions concerning individual information, is available. These models, being based on aggregated data, are also less sensitive to individual outliers. On the other hand, unit-level models take advantage of the heterogeneity of individual auxiliary information to better explain outcome variables. \citep{hidiroglou2016comparison} present a simulation study to compare area- and unit-level models, and they found that, in general, unit-level models perform better than area-level models leading to better interval coverage and more precise estimates. 

In this paper, we focus on  unit-level SAE models for the estimation of per-capita amount or rates referred to economic variables. Probability distributions of income or consumption for households and of turnover, value added and employees for firms, are typically positively skewed. Moreover, economic outcome variables, besides the heterogeneity of the location parameter, considered conditional to covariates,  generally show conditional heterogeneity also on scale, skewness and kurtosis parameters.We propose a  unit-level SAE model based on Generalized Additive Models for Location, Scale and Shape (GAMLSS). In GAMLSS several alternative distributional assumptions can be considered for the outcome variable and each distribution parameter, not only the location one, can be modelled in terms of covariates. The SAE literature almost completely focuses on the aim of modelling only the location parameter and usually relies on the Normal distributional assumption. We expect that GAMLSS peculiarities could be fruitfully exploited in the SAE prediction of parameters relating to probability distribution of economic variables, usually skewed and with high levels of individual heterogeneity.

Originally defined in the form of the nested error linear regression model (\citep{battese1988prediction1}), unit-level SAE models have been gradually generalized in more refined and flexible models (\citep{rao_small_2015}). Nowadays, to match the requests of data users asking for  more territorially detailed estimates, some features coming from  real applications have to be considered. In this perspective, dozens of different models are theorised, each one trying to overstep the assumptions of the basic unit-level linear mixed model. Some researchers propose models releasing the assumption of  Normal distribution referred to the outcome variable conditionally on covariates.  \citep{gosh1998} define a Generalized Linear Mixed Models (GLMM) where the distribution of the dependent variable is assumed to belong to the exponential family. \citep{rojas2020data} use specific transformation of the dependent variable (i.e.  Log-Shift, Box-Cox, Dual, etc.) to reach normality. Unfortunately, in the economic field, variables cannot always be brought back to normality by transformations and their distribution can often be more complex than those feasible with GLMM. Recently \citep{graf2019generalized} developed a SAE model based on the flexible Generalized Beta of the second kind (GB2) distributional assumption and \citep{lyu_empirical_2020} propose a zero-inflated Log-Normal model. In addition, non-parametric and semi-parametric models have been proposed in SAE literature. For the non-parametric we mainly refer to \citep{chambers2006m} and \citep{opsomer2008non} while for the semi-parametric a SAE Generalized Additive Mixed Model (GAMM) have been proposed by \citep{rueda2012small}. Also in the area-level context normality has been released (\citep{bell2006using}; \citep{ferrante}; \citep{hobza2023area}). Besides, in the framework of the SAE Normal models, \citep{raohetr}, \citep{jiang2012small} and \citep{heteros} deal with heteroskedasticity. 
For a more recent review of SAE methods, we refer to  \citep{tzavidis2018start}, \citep{sugasawa2020small} and \citep{rao2}.

Unit-level SAE models based on GAMLSS, at the best of our knowledge have never been considered in the SAE context. GAMLSS, introduced by \citep{GAMLSS2005}, generalize GLM and GAM, but also GAMM and GLMM models. As briefly mentioned before, some interesting features of GAMLSS can be suitably and profitably spent for the estimation of economic parameters in small areas. At first, in GAMLSS the response distribution is not restricted to belonging to the exponential family, whereas over one hundred continuous and discrete distributions can be adopted for modelling the response variable. Truncated, censored and finite mixture versions of these distributions can also be used. A family of increasingly flexible models can be defined according to the distribution that best suits the data. This opportunity is very important in the socio-economic field, where normality, although often assumed, is rare. Secondly, but perhaps even more relevant, not only the location parameter, but also each parameter of the outcome variable distribution can be modelled as functions of the covariates and random effects. This feature of GAMLSS could lead to a  further reduction of the estimates mean square error via the reduction of the bias introduced by the model. The possibility to model each parameter provides the option of borrowing strength not only by the covariates explaining the location parameter, as usual in SAE,  but also from the areal heterogeneity of the remaining parameters. 

The paper is organised as follows. Section \ref{sec2} introduces GAMLSS. In Section \ref{sec3}  we extend the GAMLSS theory to the SAE framework. The prediction of general area parameters is defined, with a parametric bootstrap proposal for MSE estimation. The proposed methods are evaluated against more common competitors using model-based  and design-based simulation studies (Section \ref{sec4}). In Section \ref{sec6} we specify a GAMLSS-SAE model for estimating, based on Household Budget Survey data, the per-capita consumption of Italian and foreigner households, in the urban and rural areas of Italian regions. Section \ref{sec7} summarises the main findings and outlines further research.

\section{Generalized Additive Models for Location, Scale and Shape}
\label{sec2}

\subsection{GAMLSS: the general definition}
GAMLSS assume independent observations   $y_i$, $i=1,\dots,n$, from a random variable \textit{Y}, with Probability Density Function (PDF)  $f(\textit{Y}|\boldsymbol{\theta}_i)$, conditional on  $\boldsymbol{\theta}_i^{T}=(\theta_{i1},\dots,\theta_{ik},\dots, \theta_{ip})$, a vector of $p$ distribution parameters, $k=1,\dots,p$. Under GAMLSS each parameter can be expressed as a function of the covariates. \citep{GAMLSS2005} define the original formulation of a GAMLSS as follows. Let $\mathbf{y}^T=(y_1,\dots,y_n)$ be the \textit{n} length vector of the response variable. Let  $g_k(\cdot)$ be a known monotonic link functions relating the \textit{p} distribution parameters to explanatory variables by:

\begin{equation}
\label{eqn:GAMLSS}
    g_k(\boldsymbol{\theta}_k)=\mathbf{X}^k\boldsymbol{\beta}_k+\sum_{m=1}^{M_k}\mathbf{Z}_{m}^k\boldsymbol{\gamma}_m^k, \quad  k=1,\dots,p \quad  \text{and} \quad m=1,\dots,M_k
\end{equation}
where $\boldsymbol{\theta}^T_k=(\theta_{1k},\dots,\theta_{nk})$ is a vector of length \textit{n}, $\boldsymbol{\beta}_k^T=(\beta_{1k}, \dots, \beta_{\rho_kk})$ is a parameter vector of length $\rho_k$, $\mathbf{X}^k$ is a known design matrix of order $n \times \rho_k$, $\mathbf{Z}_{m}^k$ is a fixed known $n \times q_{mk}$ design matrix, $\mathbf{\gamma}_m^k$ is a $q_{mk}$-dimensional random effect term and $M_k$ is the number of additive terms for each parameters. Let each parameter \textit{k} in (\ref{eqn:GAMLSS}) allow for a combination of different types and numbers of additive random effects terms, $\gamma_{m}^k$, to be incorporated easily in the model.
The distributional form of $f(\textit{Y}|\boldsymbol{\theta}_i)$  is not restricted to a specific family. Indeed, in GAMLSS, the exponential family distributional assumption, essential in GLMs and GAMs, is relaxed and replaced by a general distribution family, including highly skewed and/or kurtotic continuous and discrete distributions. 
The number of parameters \textit{p}, despite not being  constrained, is often assumed  $\le 4$ since with these limits almost every distribution could be used in the GAMLSS framework. The first two population distribution parameters are often location ($\mu$) and scale ($\sigma$) parameters, while the remaining ones, if any, are shape parameters, e.g., skewness ($\nu$) and kurtosis ($\tau$). GAMLSS allow a variety of additive smoothing terms ($\sum_{m=1}^{M_k}\mathbf{Z}_{m}^k\boldsymbol{\gamma}_m^k$) as P-spline, cubic splines, random effects, non-parametric random effects and many others. We refer to $\boldsymbol{\gamma}_m^k$ always as parametric random effects as it typically is in SAE models. 

The estimation of the parametric part of GAMLSS is achieved by either of two different algorithmic procedures, which maximise a penalized likelihood of the data (\citep{GAMLSS2005}). The first is the Newton–Raphson algorithm (RS). The second is based on the Cole and Green algorithm (CG), which is a "transformation" of the Fisher scoring algorithm. Both algorithms, for the estimation of the  distribution parameters, are performed within an outer and an inner iteration,  while a back-fitting algorithm is used to estimate $\boldsymbol{\beta }$s and $\boldsymbol{\gamma}$s parameters.
Crucial to the way that additive components are fitted within the GAMLSS framework is the back-fitting algorithm. The quadratic penalties in the likelihood derive from assuming a normally distributed random effect on the linear predictor. Assume in  (\ref{eqn:GAMLSS}) the $\boldsymbol{\gamma}_m^k$ independent normally distributed, with $\boldsymbol{\gamma}_m^k \sim N_{q_{mk}}(\mathbf{0}, [\mathbf{G}_{mk}(\boldsymbol{\lambda}_{mk})]^-)$, where $[\mathbf{G}_{mk} (\boldsymbol{\lambda}_{mk})]^-$ is the inverse of a $q_{mk}\times q_{mk}$ symmetric matrix which may depend on a vector of hyper-parameters $\boldsymbol{\lambda}_{mk}$. For fixed $\boldsymbol{\lambda}_{mk}$s the $\boldsymbol{\beta}_k$s and the $\boldsymbol{\gamma}_m^k$s are estimated by maximising the following penalized likelihood function $l_p$:
\begin{equation}
\label{likel}
    l_p=\sum_{i=1}^n\log\{f(y_i|\boldsymbol{\theta}_i)\}-\frac{1}{2}\sum_{k=1}^p\sum_{m=1}^{M_k}(\boldsymbol{\gamma}_m^k)^T\mathbf{G}_{mk}\boldsymbol{\gamma}_m^k,  \quad  i=1,\dots,n
\end{equation}
As noted by \citep{GAMLSS2005}, the maximization of  (\ref{likel}) is equivalent to using an empirical Bayesian argument to obtain the maximum a posteriori estimation
 of both the $\boldsymbol{\beta}_k$s and the $\boldsymbol{\gamma}_{m}^k$s assuming Normal, possibly improper, priors for the $\boldsymbol{\gamma}_{m}^k$s.
 
\subsection{Random effects estimation}
Define $y_{ij}$  as the observed variable for unit \textit{i} in group \textit{j}, for $i=1,\dots,n_j$ and $j=1,\dots,J$, where $n_j$ is the number of units belongings to the $j-th$ group and $n=\sum_{j=1}^J n_j$. We consider only one additive term, i.e. $M_k=1, \forall k$ and, for the moment, we take into account 
%Following \citep{GAMLSS2005} and \citep{book2017}, we assume that there are random effect variables $\boldsymbol{\gamma}$  providing an additional source of variation, which have PDF $f(\boldsymbol{\gamma}|\boldsymbol{\lambda})$. Under GAMLSS, in  (\ref{eqn:GAMLSS}), the response variables $Y$ are assumed to be independently distributed
%with PDF  $f(\text{Y}|\boldsymbol{\beta},\boldsymbol{\gamma})$. This assumption holds that: 
%\begin{equation}
%\label{distraneff}
%    \underbrace{f(\text{Y}|\boldsymbol{\beta},\boldsymbol{\lambda})}_\text{marginal Y} =\int_{\boldsymbol{\gamma}}
%      \overbrace{\underbrace{f(\text{Y}|\boldsymbol{\beta},\boldsymbol{\gamma})}_\text{conditional $\text{Y}|\boldsymbol{\gamma}$}\quad \underbrace{f(\boldsymbol{\gamma}|\boldsymbol{\lambda})}_\text{marginal $\boldsymbol{\gamma}$}}^\text{joint Y and $\boldsymbol{\gamma}$} d\boldsymbol{\gamma}
%\end{equation}
%where, given $\boldsymbol{\beta}$ and $\boldsymbol{\lambda}$,  $f(\text{Y}|\boldsymbol{\beta},\boldsymbol{\lambda})$ denotes the marginal distribution of $Y$.  $f(\text{Y}|\boldsymbol{\beta},\boldsymbol{\gamma})$ is the conditional distribution of $Y$, given $\boldsymbol{\gamma}$ and $f(\boldsymbol{\gamma}|\text{Y})$ is the marginal distribution for the random effects $\boldsymbol{\gamma}$. 
a model with four parameters $(\boldsymbol{\mu}, \boldsymbol{\sigma}, \boldsymbol{\nu}, \boldsymbol{\tau})$ where only $\boldsymbol{\mu}$ is defined in terms of random effects. As noted by \citep{GAMLSS2005}, when random effects are considered, the matrix  $\mathbf{Z}$ is defined as an identity matrix $\mathbb{I}_n$. This case is easily extendable to those where the random effects are defined for more than one parameters (\citep{GAMLSS2019}, Ch. 10). Re-writing (\ref{eqn:GAMLSS}) as: 
\begin{equation}
\label{mt}
    \begin{cases}
    g_\mu(\boldsymbol{\mu}_{ij})=\mathbf{X}^\mu_{ ij}\boldsymbol{\beta}^\mu+\mathbf{Z}^\mu_j\boldsymbol{\gamma}_{j}^\mu\\
    g_\sigma(\boldsymbol{\sigma}_{ij})=\mathbf{X}^\sigma _{ij}\boldsymbol{\beta}^\sigma\\
    g_\nu(\boldsymbol{\nu}_{ij})=\mathbf{X}^\nu _{ij}\boldsymbol{\beta}^\nu\\
    g_\tau(\boldsymbol{\tau}_{ij})=\mathbf{X}^\tau _{ij}\boldsymbol{\beta}^\tau\\
    \end{cases}
\end{equation}
where $\boldsymbol{\gamma}^T=(\gamma_1,\dots, \gamma_j \dots,\gamma_J)$. Assume now that: 
\begin{equation}
    \gamma_j\sim N(0,\sigma^2_\gamma) \quad \text{independently for} \quad j=1,\dots,J
    \label{3eq}
\end{equation}
The model defined by (\ref{mt}) and (\ref{3eq}) may be fitted using Gaussian quadrature to approximate the integral of the equation. Following \citep{GAMLSS2019}, from (\ref{3eq}) the vector $\boldsymbol{\gamma}$ has a Normal distribution: 
\begin{equation*}
    \boldsymbol{\gamma}\sim N(\textbf{0},\sigma^2_{\boldsymbol{\gamma}} \mathbb{I})
\end{equation*}
Defining $\gamma_j=\sigma_{\boldsymbol{\gamma}} U_j$ where $U_j\overset{ind}{\sim}N(0,1)$, follows that: 
\begin{equation}
    \mathbf{U}=\frac{\boldsymbol{\gamma}}{\sigma_{\boldsymbol{\gamma}}}\sim N(\textbf{0},\textbf{1})
\end{equation}
where $\mathbf{U}^T=(U_1,\dots,U_J)$. Gaussian quadrature effectively approximates the continuous $N(0, 1)$ distribution for each $U_j$ by a discrete distribution: 
\begin{equation}
   \label{guass}
   Prob(U_j=u_\kappa)=\pi_\kappa, \quad \kappa=1,\dots,\mathcal{K}
\end{equation}
where the $u_\kappa$s and the $\pi_\kappa$s are fixed and known for a fixed total number of discrete points $\mathcal{K}$ used for the Gaussian quadrature approximation. The model (\ref{guass}) can be now considered as a finite mixture of $\mathcal{K}$ components in which the prior probability $\pi_\kappa$s are fixed and, once the total number
of quadrature points $\mathcal{K}$ has been chosen, the $u_\kappa$s are also fixed and known. Hence, the predictor for $\mu_{ij\kappa}$ is: 
\begin{equation}
    g_\mu(\boldsymbol{\mu}_{ij\kappa})=(\mathbf{X}^{\mu}_{ ij})^T\boldsymbol{\beta}^\mu +\sigma_\gamma u_\kappa
\end{equation}
with prior probability $\pi_\kappa$. This model is estimable with an EM algorithm (\citep{GAMLSS2019}, Ch. 10). 
\subsection{The inferential framework }
\label{sec:inf}
Here we highlight some inferential aspects related to GAMLSS. In order to answer the classical inference question, e.g. model fitting, testing values and, model selection, \citep{GAMLSS2005} compare likelihood-based inference and bootstrapping. \citep{kneib2013beyond} analyses the Bayesian inference and functional gradient descent boosting while \citep{aeberhard2021robust} propose the use of robust selection of the smoothing parameters for GAMLSS. Moreover, \citep{wood2020inference} and \citep{kneib2020comments} present a review of the inferential framework for additive regression models with smoothers and/or random effects including GAMLSS.

In our view, the model selection in the GAMLSS framework  deserves further attention. Let $\mathcal{M}=\{\mathcal{D},\mathcal{G},\mathcal{T},\mathcal{L}\}$ represent a GAMLSS as defined in (\ref{eqn:GAMLSS}), where the components of $\mathcal{M}$ represent: $\mathcal{D}$ the set of the possible distribution of the response variable, $\mathcal{G}$ the set of link functions, $\mathcal{T}$ the set of the terms appearing in all the predictors and $\mathcal{L}$ the set of the smoothing hyper-parameters. In GAMLSS the model selection operates starting from $\mathcal{D}$ till $\mathcal{L}$. Here, we focus only on the choice within the sets $\mathcal{D}$ and $\mathcal{T}$ that are peculiar for SAE (for the selection referred to the other components see \citep{GAMLSS2019}, Ch. 11).
For a parametric statistical model $\mathcal{M}$ and within a likelihood-based inferential procedure, the distribution $\mathcal{D}$ is selected by the global deviance $GDEV=-2l(\widehat{\boldsymbol{\theta}})$. Let  $\mathcal{M}_0$ and  $\mathcal{M}_1$ be nested statistical models with fitted global deviances $GDEV_0$ and $GDEV_1$ and degrees of freedom $df_0$ and $df_1$, respectively. $\mathcal{M}_0$ and $\mathcal{M}_1$ may be compared using the generalized likelihood ratio test statistic which, under certain conditions, has asymptotically the $\chi^2_d$ distribution where $d=df_0-df_1$.  To compare non-nested GAMLSS models, the generalized Akaike information criterion (GAIC) could be used to penalize over-fitting. This is obtained by adding to the fitted deviance a penalty for each effective degree of freedom used in the model. \citep{GAMLSS2005} suggest starting with a penalty equal to $\sqrt{\log(n)}$ if $n\ge 1000$ and to  use different values of the penalty with the aim of investigating the sensitivity or robustness of model selection. In short, given that GAMLSS, unlike the classical regression models, consider more than one parameter to be explained by covariates, the selection of $\mathcal{D}$ is done using GDEV instead of the deviance and there is the necessity to use a specific penalty. The selection within the set $\mathcal{T}$ requires specific procedures that could be seen as an extension and generalisation of the usual ones. There are four possible procedures: (i) the criterion-based methods, based on the GAIC criterion (\citep{GAMLSS2019}, Ch. 11), (ii) the regularization method, which relies on the ridge and lasso regression concept and which can also be extended to the GAMLSS framework, (iii) the boosting algorithm discussed by \citep{hofner2014gamboostlss}, emerging from the field of supervised machine learning and, 
nowadays, often applied as a flexible alternative to estimate and select predictor effects,  (iv) the dimension-reduction method, which uses a principal component analysis to choose the covariates. In particular, focusing on the criterion-based methods, \citep{GAMLSS2005} propose two different strategies: (i) the "fixed order procedure" in which the selection take place in the following order: first $\mu$ followed by $\sigma$, then $\nu$ and finally $\tau$, (ii) the "fixed terms procedure" which forces all the distribution parameters to have the same terms, then it is not possible to use different predictors depending on the parameters. 

%Before to move on, let us note, as also the diagnostic for GAMLSS differ from the one defined for the other regression's methods. Indeed, for GAMLSS the diagnostic is based on normalized randomized quantile residuals which are derived thanks to the the probability integral transform. For the proof and for a more complete explanation about the inadequacy of the classical residuals see \citep{GAMLSS2019}. 

%for more on the proof and the reasons 

%As noted by \citep{GAMLSS2019} the raw residuals (simple linear regression) are difficult to generalize to distributions other than the normal, while the  
%the deviance residuals and the Pearson residuals (both used for the GLM) have problems if the modelled parameters are more than one and/or response variable is highly skewed or kurtotic. \citep{GAMLSS2019} propose to use normalized randomized quantile residuals which are derived thanks to the the probability integral transform. In addition, in GAMLSS, the worm plot, a diagnostic tool for checking the residuals within different ranges of the explanatory variables, was proposed to identify intervals of the explanatory variable within which the model does not adequately fit the data. 

\section{GAMLSS for Small Area Estimation} 
\label{sec3}
In this section, we propose a SAE model based on GAMLSS. We focus at first on the specification and estimation of the model and then on the prediction of the individual values of the outcome variable for the non sampled part. In small area estimation, a finite population $\Omega$ with \textit{N} units is distributed into \textit{J} sub-population $\Omega_1, \dots, \Omega_J$, called domains or areas of size $N_1, \dots, N_J$, where $N=\sum_{j=1}^J N_j$. We identify with $Y_{ij}$ the target variable and with $y_{ij}$ the observation, for $i=1,\dots,n$ and $j=1,\dots,J$. Following the notation used from equation (\ref{eqn:GAMLSS}) on-wards, the area specific random effects are defined as $\mathbf{Z}_{j}^k\gamma_j^k$.
 SAE models aim to estimate area parameters in the form of $H_j=\zeta(Y_{ij})$ where $\zeta(\cdot)$ is a real measurable function. Design-based estimates of these area parameters are obtained using data from the sample \textit{s} of size \textit{n} drawn from the population $\Omega$ with sub-sample $s_j =s \cap \Omega_j$ for area \textit{j} of size $n_j$, being $n=\sum_{j=1}^J n_j$. We denote by $r_j=\Omega_j-s_j$ the sample complement from the area $j$. Small area models need to adopt a strategy allowing specific area variation. In this context, mixed models are particularly interesting as they involve random area specific effects, which make it possible to add between areas heterogeneity to that introduced by covariates. We specify a GAMLSS small area model, starting from  (\ref{eqn:GAMLSS}), by considering area specific random effects and assuming $Z \equiv \mathbb{I}_n$. Moreover, we limit our attention to four parameter distributions.
\begin{equation}
\label{m2}
    \begin{cases}
    g_\mu(\boldsymbol{\mu}_{ij})=\mathbf{X}^\mu_{ ij}\boldsymbol{\beta}^\mu+\boldsymbol{\gamma}_j^{\mu}\\
    g_\sigma(\boldsymbol{\sigma}_{ij})=\mathbf{X}^\sigma_{ ij}\boldsymbol{\beta}^\sigma+\boldsymbol{\gamma}_j^{\sigma}\\
    g_\nu(\boldsymbol{\nu}_{ij})=\mathbf{X}^\nu_{ ij}\boldsymbol{\beta}^\nu+\boldsymbol{\gamma}_j^{\nu}\\
    g_\tau(\boldsymbol{\tau}_{ij})=\mathbf{X}^\tau_{ ij}\boldsymbol{\beta}^\tau+\boldsymbol{\gamma}_j^{\tau}
    \end{cases}
\end{equation}
In (\ref{m2}) $\boldsymbol{\gamma}_j^k\stackrel{iid} \sim N(0, \boldsymbol{\Psi}_k)$ for $k=\mu,\sigma,\nu,\tau$, allows to consider differences among areas for each parameter. Random effects are assumed to be independent (\citep{GAMLSS2019}, Ch. 10). The variance-covariance matrix $\boldsymbol{\Psi}_k$	involves the variance of the random effects $\sigma^2_k$, for independent random effects we have $\boldsymbol{\Psi}_k=\sigma^2_k\mathbb{I}$. 
 In (\ref{m2}) the random effects referred to $\boldsymbol{\mu}, \boldsymbol{\sigma}, \boldsymbol{\nu}$ and $\boldsymbol{\tau}$ are assumed: 
\begin{equation*}
    \boldsymbol{\gamma}_j^\mu\sim N(0,\sigma^2_\mu), \quad \boldsymbol{\gamma}_j^\sigma\sim N(0,\sigma^2_\sigma), \quad \boldsymbol{\gamma}_j^\nu\sim N(0,\sigma^2_\nu) \quad \text{and} \quad \boldsymbol{\gamma}_j^\tau\sim N(0,\sigma^2_\tau)
\end{equation*}
The distribution model $\mathcal{F}$ allows for both skewness and kurtosis in the conditional distribution of $Y_{ij}$: 

\begin{equation}
\label{distr}
    Y_{ij}|\gamma_j^{\mu},\gamma_j^{\sigma},\gamma_j^{\nu},\gamma_j^{\tau}\sim \mathcal{F}(Y_{ij}|\mu_{ij},\sigma_{ij},\nu_{ij},\tau_{ij}),  \quad i=1,\dots,n_j \quad \text{and} \quad j=1,\dots,J
\end{equation}
Assuming (\ref{distr}) as the correct model for the population, the estimates of the parameters can be obtained by one of the methods mentioned in Section \ref{sec2}. Estimates  lead to $\mathcal{F}(y_{ij}|\widehat{\mu}_{ij},\widehat{\sigma}_{ij},\widehat{\nu}_{ij},\widehat{\tau}_{ij}), \quad i\in s \quad \text{and} \quad j=1,\dots,J$ where $i \in s$ indicate that the parameters are estimated using the sample part of the population. Note as the sample weights are not used in the estimation of the model (\ref{m2}) as usually done in the unit-level SAE models. In the next section, without loss of generality, we assume a continuous dependent variable. What follows could be generalized with some mathematical precautions also to discrete distributions.

\subsection{Prediction of general small area parameters}
We propose a strategy to estimate a general small area parameter $H_j=\zeta(Y_{ij})$ through the estimator $\tilde{H}_j=E[\zeta(Y_{ij})]$. Denote with $\widehat{H}_j$ the prediction of $H_j$. Note that the random variable $Y_{ij}$ could be split in two non-overlapped sets $y_{ij}^s$ and $Y_{ij}^r$, where $y_{ij}^s$ refer to the sets of the observed sample units and $Y_{ij}^r$ to the set of the non sample units. %Under model (\ref{distr}) the quantity  $Y_{ij}^r$  is still a random variable for which we assume that hold that the estimated values of $\mu, \sigma, \nu, \tau$ and  $\gamma$, obtained from the sample part of the population. %, hold also for the non sample part. 
In the model based approach, the estimated value $\widehat{H}_j$ can be expressed as weighted mean of $\widehat{H}_j^r$ and  $H_j^s$, where $\widehat{H}_j^r$ is the estimated parameter for the non sample part of the population and  $H_j^s$ is the known observed quantity for the sample part of the population. The estimator of $H_j^r$ referred to the  $j-th$ area can be expressed as follows:
%If the (\ref{distr}) is known, the estimator of $H_j$ is expressed by: 
%\begin{equation}
%\label{intGA}
%    \Tilde{H}_j=E[\zeta(Y_{ij})]=\int_\Re \zeta(Y_{ij})f(Y_{ij}|\mu_{ij},\sigma_{ij},\nu_{ij},\tau_{ij})dY_{i}, \quad j=1,\dots,J
%\end{equation}
% Through the sample part, is possible to properly estimate $\mu, \sigma, \nu, \tau$ and  $\gamma$ obtaining $\widehat{\delta}^T=(\widehat{\gamma}, \widehat{\mu},\widehat{\sigma},\widehat{\nu},\widehat{\tau})$. Where $\widehat{\gamma}, \widehat{\mu},\widehat{\sigma}$ and  $\widehat{\nu}$ are function of $\widehat{\beta}$.  Under the assumption that the estimated values obtained from the sample part of the population hold also for the non sample part we can estimate 
% outcome small area parameter $\widehat{H}_j^r$ can be estimated, as follows:
\begin{equation}
\label{intgans}
    \tilde{H}_j^r=E[\zeta(Y^r_{ij})]=\int_{\Re^{N_j-n_j}} \zeta(Y^r_{ij})f(Y^r_{ij}|\hat{\mu}_{ij},\hat{\sigma}_{ij},\hat{\nu}_{ij},\hat{\tau}_{ij})dY^r_{i}, \quad i \in r \quad \text{and} \quad j=1,\dots,J
\end{equation}
where $f(Y^r_{ij}|\hat{\mu}_{ij},\hat{\sigma}_{ij},\hat{\nu}_{ij},\hat{\tau}_{ij})$ is the PDF for the non sampled units in which parameters are assumed to be the same as for the sample part. Computing the integral in (\ref{intgans}) may be cumbersome. Denoting the vector of estimated values of $\mu, \sigma, \nu, \tau$ and  $\gamma$ as $\widehat{\delta}^T=(\widehat{\gamma}, \widehat{\mu},\widehat{\sigma},\widehat{\nu},\widehat{\tau})$,  we use a Monte Carlo (MC) procedure to approximate $\widehat{H}_j$: 
\begin{enumerate}
\label{MCapprox}
    \item fit the $\mathcal{F}$ model (\ref{distr}) to the sample data, obtaining a consistent estimate $\hat{\delta}$ of $\delta$; 
    \item for each $\ell =1,\dots,L$, for \textit{L} large, generate an out-of-sample vector ${}^{(\ell)}\mathbf{y}^{r}_{ij}$ from: 

\begin{equation*}
 \mathcal{F}(Y^r_{ij}|\hat{\mu}_{ij},\hat{\sigma}_{ij},\hat{\nu}_{ij},\hat{\tau}_{ij}) \quad\text {with} \quad i \in r
\end{equation*}
    \item attach the sample data $\mathbf{y}_{ij}^s$ to the generated out-of-sample data ${}^{(\ell)}\mathbf{y}^{r}_{ij}$ to form the area population vector ${}^{(\ell)}\mathbf{y}_{ij}^T=(({}^{(\ell)}\mathbf{y}^{r}_{ij})^T, (\mathbf{y}^{s}_{ij})^T)$. With  ${}^{(\ell)}\mathbf{y}_{ij}^T$ is possible to calculate the target parameter\\ $H^{(\ell)}_j=\zeta({}^{(\ell)}\mathbf{y}_{ij})$. 
    \end{enumerate}
A MC approximation of $\widehat{H}_j$ is then: 
    \begin{equation*}
        \widehat{H}_j \approx \frac{1}{L}\sum_{i =1}^LH^{(\ell)}_j
    \end{equation*}
 \noindent   
The procedure described is useful when the specific parameters we are estimating do not have closed form expressions. Below, we provide the closed-form expressions for the predictors of the area means. 
Under model (\ref{distr}) the predictor of the area means $\bar{Y}_j=N^{-1}_j\sum_{i=1}^{N_j}Y_{ij}$ is given by: 
\begin{equation}
    \widehat{\bar{Y}_j}=N_j^{-1}\Biggl(\sum_{i\in s_j}y_{ij}+\sum_{i\in r_j}\widehat{y}_{ij}^r\Biggr)
\end{equation}
where $\widehat{y}_{ij}^r$ is the predicted value for the non sample unit $i \in r_j$. Note that, if $\mathcal{F}$ is a distribution with the mean expressed as a linear function of $\mu$ only and $g_\mu(\cdot)$ is the identity link function, we have:
    \begin{equation}
    \label{Normgam}
        \widehat{\bar{y}}^r_{j}=\int_{\Re^{N_j-n_j}} \widehat{y}_{ij}^r f(Y^r_{ij}|\widehat{\mu}_{ij}, \widehat{\sigma}_{ij}, \widehat{\nu}_{ij}, \widehat{\tau}_{ij})dY^r_{i}=\widehat{\beta} \bar{X}_j+\mathbf{Z}_j\widehat{\gamma}_{j}, \quad i \in r 
    \end{equation}
    where $\bar{X}_j$ is the matrix of the area mean of the covariates and  (\ref{Normgam}) is exactly the same formulation of the classical EBLUP with the assumption of normality. 

\subsection{Parametric bootstrap for MSE estimation}
\label{sec:MSE}
The mean square error (MSE) estimation is one of the main points in the SAE framework. Here we suggest a bootstrap procedure to estimate it. The expected value of the squared difference between the estimated value $\widehat{H}_j$ and its real value counterpart $H_j$ is: 
\begin{equation}
\label{MSEDIM}
    MSE_j=E[(\widehat{H}_j- H_j)^2] \quad \text{for} \quad j=1,\dots, J
\end{equation}
Obviously, (\ref{MSEDIM}) is not calculable because it requires the knowledge of the true parameters. Moreover, a unique analytic approximation of the MSE estimator is unfeasible given that we consider a generic area parameter $H_j$ and a generic link function $g_k(\cdot)$. A specific analytic formulation can only be found by specifying an area parameter and a link function. For example, the analytic MSE estimator assuming a logistic model to estimate the mean of the areas will be the same as \citep{gonzalez2007estimation}. An estimator of the MSE of a GAMLSS's $\widehat{H}_j$ can be obtained using the parametric bootstrap for finite populations. This method proceeds as follows:
\begin{enumerate}
    \item fit the unit-level model with one of the allowed algorithms to obtain model parameter estimators  $\widehat{\boldsymbol{\delta}}$ and $\widehat{\boldsymbol{\Psi}}_k$ for $k=\mu,\sigma,\nu,\tau$;
    \item given the estimates obtained in step 1, for each $k$ generate a vector $\mathbf{t}^\ast_k$ whose elements are \textit{J} independent realisations of a  \textit{N(0,1)} variable. For each parameter  \textit{k}, construct the bootstrap vector $\boldsymbol{\gamma}^{k\ast}=\widehat{\boldsymbol{\Psi}}_k\mathbf{t}_k^\ast$ and generate:
    \begin{equation*}
    \begin{cases}
    \widehat{\boldsymbol{\mu}}_{ij}^\ast=g_\mu^{-1}(\mathbf{X}^\mu_{ ij}\widehat{\boldsymbol{\beta}}^\mu+\boldsymbol{\gamma}_j^{\mu\ast})\\
    \widehat{\boldsymbol{\sigma}}_{ij}^\ast=g_\sigma^{-1}(\mathbf{X}^\sigma_{ ij}\widehat{\boldsymbol{\beta}}^\sigma+\boldsymbol{\gamma}_j^{\sigma\ast})\\
    \widehat{\boldsymbol{\nu}}_{ij}^\ast=g_\nu^{-1}(\mathbf{X}^\nu_{ ij}\widehat{\boldsymbol{\beta}}^\nu+\boldsymbol{\gamma}_j^{\nu\ast})\\
    \widehat{\boldsymbol{\tau}}_{ij}^\ast=g_\tau^{-1}(\mathbf{X}^\tau_{ ij}\widehat{\boldsymbol{\beta}}^\tau+\boldsymbol{\gamma}_j^{\tau\ast})
    \end{cases}
\end{equation*}
    \item for each \textit{j} and each \textit{k} generate a bootstrap population of response variables from the  model $\mathcal{F}(\hat{\mu}^\ast_{ij},\hat{\sigma}^\ast_{ij},\hat{\nu}^\ast_{ij},\hat{\tau}^\ast_{ij})$;  
    \item let $\mathbf{Y}_{ij}^{{P*}^T}=(y_{j1}^{\ast},\dots,y_{jN_j}^{P*})$  denote the vector of generated bootstrap response variables for area  \textit{j}. Calculate target quantities for the bootstrap population as $H_j^{\ast}=\zeta(\mathbf{Y}_{ij}^{P*})$;
    \item let $\mathbf{y}_s^{\ast}$ be the vector whose elements are the generated $y_{ij}^{\ast}$ with indices contained in the sample  \textit{s}. Fit the model to the bootstrap sample data $\{(y_{ij}^{\ast},\mathbf{X}_{ij}; i\in s)\}$ 
and obtain the bootstrap model parameter estimators;
\item obtain the bootstrap estimator of $H_j$ through the Monte Carlo approximation, denoted $\widehat{H}^{\ast}_j$;
\item repeat steps [2-6] a large number of times  \textit{B}. Let $H_j^{\ast}(b)$ be the true value and  $\widehat{H}_j^{\ast}(b)$ the estimator obtained in $b-th$ replicate of the bootstrap procedure, $b=1,\dots,B$.
\end{enumerate}
The bootstrap mean square-error estimator of $\widehat{H}_j^{}$ is given by: 
\begin{equation}
\label{eq:mseb}
    \widehat{MSE}_B(\widehat{H}_j)=B^{-1}\sum_{b=1}^B\biggl[\widehat{H}_j^{\ast}(b)- {H}_j^{\ast}(b)\biggr]^2
\end{equation}

\section{Model- and design-based simulations}
\label{sec4}
We study the characteristics and performance of small area predictors based on GAMLSS under different scenarios with model- and design-based simulations. We comparatively evaluate the performance of the GAMLSS estimator and of the well-known unit-level Battese-Harter-Fuller (BHF, \citep{battese1988prediction1}). We refer to BHF estimator as EBLUP.  When the dependent variable is not Normal we use a BHF with Box-Cox transformation (\citep{rojas2020data}). We consider as parameter of interest averages or proportions which, for income and consumption variables, could be per-capita income or consumption, poverty rates, budget shares, and so on. For every simulation we define a different distribution model $\mathcal{F}$ using, for each parameter of the chosen distribution, the link function that minimize the GAIC and a set of different covariates. The code used for the simulations is public available on GitHub: \textit{saegamlss}. 

\subsection{Model-based simulations}

Model-based (MB) simulation is frequently used to determine estimator characteristics when it is hard to achieve analytic results on estimator properties.  We perform simulations based on four different models and, eventually, different outcome parameters:  
 \begin{itemize}
     \item [\textit{(A)}] \textit{MB Normal}: Normal model for the estimation of the mean. A GAMLSS, where normality is assumed and only $\mu$ is defined in terms of covariates, is equivalent to a linear mixed-effects model. In this simulation the model is defined as: 
     \begin{equation*}
         \boldsymbol{\mu}_{ij}=\beta_0^\mu+\beta_1^\mu\mathbf{X}_{1ij}+\boldsymbol{\gamma}^\mu_j+\boldsymbol{\epsilon}_{ij}
     \end{equation*}
     \noindent
     Data are generated by assuming $\beta_0^\mu=100$ and $\beta_1^\mu=4$ while the value of the variance of $\boldsymbol{\gamma}$ and $\boldsymbol{\epsilon}$ changes from one scenario to another as follows: 
\begin{equation*}
\begin{split}
\textbf{Scenario 1} \quad \boldsymbol{\gamma}_j^\mu\sim N(0,4^2) \quad \text{and}\quad \boldsymbol{\epsilon}_{ij} \sim N(0,20^2)\\
\textbf{Scenario 2} \quad \boldsymbol{\gamma}_j^\mu\sim N(0,6^2) \quad \text{and}\quad \boldsymbol{\epsilon}_{ij} \sim N(0,22^2)\\
\textbf{Scenario 3} \quad \boldsymbol{\gamma}_j^\mu\sim N(0,8^2) \quad \text{and}\quad \boldsymbol{\epsilon}_{ij} \sim N(0,24^2)
\end{split}
\end{equation*}
We compare GAMLSS estimators obtained under Normal distribution with EBLUP estimators expecting the same results, both referring to estimates and relative MSE.

     \item [\textit{(B)}] \textit{\textit{MB Heteroskedastic Normal}}: Normal model for the estimation of the mean with heteroskedastic data. In GAMLSS both location and scale are modelled depending on covariates. We expect that the estimator could reduce the MSE compared with EBLUP MSE since, EBLUP in presence of heteroskedasticity is unbiased but inefficient and lead to biased estimates of standard errors. As well summarised by \citep{heteros}, how to handle heteroskedasticity in SAE is still an active field of research.  Our interest here is to verify if this model is able to reduce the CVs and if the MSE estimator we propose in Section \ref{sec:MSE} is able to catch the real path of the MSE. In this simulation we use two covariates, $\textbf{X}_{1}$ and $\textbf{X}_{2}$, where the $\textbf{X}_{2}$ is used to add heteroskedasticity to our data. The GAMLSS predictor is based on a Normal distribution with: 
        \begin{equation*}
            \begin{cases}
            \boldsymbol{\mu}_{ij}= \beta_0^\mu+\beta_1^\mu \textbf{X}_{1ij}+\beta_2^\mu \textbf{X}_{2ij}+\boldsymbol{\gamma}_j^\mu+\boldsymbol{\epsilon}_{ij}\\
           \boldsymbol{\sigma}_{ij}=\exp(\beta_0^\sigma+\beta_1^\sigma \textbf{X}_{2ij}+\boldsymbol{\gamma}_j^\sigma)
            \end{cases}
        \end{equation*}
The data generation process we use is different and is based on \begin{equation*}
    \boldsymbol{\mu}_{ij}=100+10\textbf{X}_{1ij}+8\textbf{X}_{2ij}+\boldsymbol{\gamma}^\mu_j+\textbf{w}_{ij}\boldsymbol{\epsilon}_{ij}
\end{equation*} The term $\textbf{w}_{ij}$, useful to reproduce heteroskedasticity ( \citep{ramirez2021random}), is defined in terms of unit and area-level components as follows: $\textbf{w}_{ij}=\boldsymbol{\gamma}^\sigma_j+0.1\textbf{X}_{2ij}$ with $\boldsymbol{\gamma}_j^\mu\sim N(0,6)$ and $\boldsymbol{\epsilon}_{ij} \sim N(0,22)$ while the value of the variance of $\boldsymbol{\gamma}_j^\sigma$ changes from one scenario to another as follows: 
\begin{equation*}
\begin{split}
&\textbf{Scenario 1} \quad \boldsymbol{\gamma}_j^\sigma\sim N(0,0.8^2) \\
&\textbf{Scenario 2} \quad \boldsymbol{\gamma}_j^\sigma\sim N(0,1^2) \\
&\textbf{Scenario 3} \quad \boldsymbol{\gamma}_j^\sigma\sim N(0,1.2^2) 
\end{split}
\end{equation*}
We compare GAMLSS with a classical EBLUP  in order to verify if our model reduces the overestimation of the MSE typical of the EBLUP. 

     \item [\textit{(C)}] \textit{MB Log-Normal}: Log-Normal model, commonly adopted in  literature to model consumption expenditure (\citep{battistin2009consumption}). We focus on the estimation of the mean where both location and scale parameters are modelled depending on covariates. Note that in SAE, positively skewed variables are usually modelled through a Normal model with transformation leading the variable to normality. The model is: 
     \begin{equation*}
     \begin{cases}
    \boldsymbol{\mu}_{ij}=exp(\beta_0^\mu+\beta_1^\mu\textbf{X}_{1ij}+\boldsymbol{\gamma}_j^\mu) \\
      \boldsymbol{\sigma}_{ij}=exp(\beta_0^\sigma+\beta_1^\sigma \textbf{X}_{2ij}+\boldsymbol{\gamma}_j^\sigma)
     \end{cases}
     \end{equation*}
      
Data are generated assuming $\beta_0^\mu=7$, $\beta_1^\mu=1$, $\beta_0^\sigma=-2$ and $\beta_1^\sigma=0.5$, the random effects are sampled by: 
\begin{equation*}
\begin{split}
&\textbf{Scenario 1} \quad \boldsymbol{\gamma}_j^\mu\sim N(0,0.6^2) \quad \text{and}\quad \boldsymbol{\gamma}_j^\sigma \sim N(0,0.2^2)\\
&\textbf{Scenario 2} \quad \boldsymbol{\gamma}_j^\mu\sim N(0,0.4^2) \quad \text{and}\quad \boldsymbol{\gamma}_j^\sigma \sim N(0,0.3^2)\\
&\textbf{Scenario 3} \quad \boldsymbol{\gamma}_j^\mu\sim N(0,0.6^2) \quad \text{and}\quad \boldsymbol{\gamma}_j^\sigma \sim N(0,0.4^2)
\end{split}
\end{equation*}
We compare GAMLSS estimators obtained under Log-Normal distribution with Box-Cox EBLUP estimators.
     \item [\textit{(D)}] \textit{MB Dagum}: Dagum model, commonly used to model income distribution. We focus on the estimation of the poverty rate by modelling all three Dagum  distribution parameters depending on covariates. In this simulation, unlike the previous ones, in the three scenarios not only the value of the variance of the random effects changes, but also the definition of $\boldsymbol{\sigma}$ and $\boldsymbol{\nu}$, as follows: 
     
\begin{align*}
\textbf{Scenario 1} \begin{cases}
\boldsymbol{\mu}_{ij}=exp(\beta_0^\mu+\beta_1^\mu\textbf{X}_{1ij}+\boldsymbol{\gamma}_j^\mu), \quad \text{with} \quad \boldsymbol{\gamma}_j^\mu\sim N(0,0.15^2)\\
 \sigma_{ij}=3.4\\
 \boldsymbol{\nu}_{ij}=exp(\beta_0^\nu+\beta_1^\nu\textbf{X}_{2ij})
\end{cases}\\
\textbf{Scenario 2} \begin{cases}
\boldsymbol{\mu}_{ij}=exp(\beta_0^\mu+\beta_1^\mu \textbf{X}_{1ij}+\boldsymbol{\gamma}_j^\mu), \quad  \text{with} \quad \boldsymbol{\gamma}_j^\mu\sim N(0,0.15^2)\\
\boldsymbol{\sigma}_{ij}=exp(\beta_0^\sigma+\beta_1^\sigma\textbf{X}_{2ij}+\boldsymbol{\gamma}_j^\sigma), \quad \text{with} \quad \boldsymbol{\gamma}_j^\sigma \sim N(0,0.1^2)\\
\boldsymbol{\nu}_{ij}=0.6\\
\end{cases}\\ 
\textbf{Scenario 3}\begin{cases}
\boldsymbol{\mu}_{ij}=exp(\beta_0^\mu+\beta_1^\mu\textbf{X}_{1ij}+\boldsymbol{\gamma}_j^\mu), \quad  \text{with} \quad \boldsymbol{\gamma}_j^\mu\sim N(0,0.15^2)\\
\boldsymbol{\sigma}_{ij}=exp(\beta_0^\sigma+\beta_1^\sigma\textbf{X}_{2ij}+\boldsymbol{\gamma}_j^\sigma), \quad  \text{with} \quad \boldsymbol{\gamma}_j^\sigma \sim N(0,0.1^2)\\
\boldsymbol{\nu}_{ij}=exp(\beta_0^\nu+\beta_1^\nu\textbf{X}_{2ij})
\end{cases}
\end{align*}
\noindent
Data are generated assuming $\beta_0^\mu=3$, $\beta_1^\mu=1.5$, $\beta_0^\sigma=1.2$, $\beta_1^\sigma=0.1$, $\beta_0^\nu=-0.4$ and $\beta_1^\nu=0.1$. We compare GAMLSS estimators obtained under Dagum distribution with Box-Cox EBLUP estimators.
\end{itemize}

 In the following Section we refer to the different simulation framework  as \textit{\textit{(A)  MB Normal}}, \textit{(B) \textit{MB Heteroskedastic Normal }}, \textit{(C) MB Log-Normal} and \textit{\textit{(D) MB Dagum}}.  In all simulations the population $\Omega$ is fixed to 50,000 with a number of areas $J=50$ and a population of size $N_j=1000$. The $n_j$ units, where $n_j$ varies among areas, are sampled based on a stratified random design without replacement. The sample size goes from a minimum of 4 units to a maximum of 61. In all scenarios the covariates are kept fixed and sampled from standard Normal distribution (\citep{liu2022empirical}). All the models are chosen to reproduce realistic situations either in the type of distribution or in the single coefficients. They are chosen in order to reproduce the correspondent estimates, i.e. the Log-Normal parameters values are based on those estimated for the Italian household consumption expenditure. The GAMLSS are estimated with a mixed algorithm with a number of loops equal to $N_{RS}=250$ and $N_{CG}=250$ and the bootstrap for MSE estimations is performed with $B=200$ iterations. Each simulation is performed with a number of loop T=500. The other parameters, which have to be fixed to perform the algorithms, are reported in the appropriate sections.

 \subsection{Design-based simulations}
 Design-based (DB) simulations are carried out to analyse the properties of the estimators when the data generation process is unknown. In this case, two different economic surveys play the role of a pseudo-population that is kept fixed: the Italian Household Budget Survey (HBS) and the Italian Statistics on Income and Living Conditions (IT-SILC). For realistic simulations we consider only covariates known in population. Moreover, for each model we select only significant covariates. For model selection we decide to use the "fixed order procedure" (Section \ref{sec:inf}) to have the possibility to use to use different covariates for different parameters. For EBLUP a classical Akaike step-wise procedure is used. The two simulations assume respectively: 
\begin{itemize}
    \item [\textit{(A)}] \textit{DB consumption data}:  in this simulation we refer to data collected in 2019 HBS. This survey, conducted by the Italian Statistical Institute (ISTAT), is based on a sample of approximately 18,000 households, which is representative of all Italian households. Data are collected on the basis of a two-stage sample design where the first stage refers to municipalities and the second stage to households.
We target the per-capita expenditure based on equivalent consumption obtained through the Carbonaro equivalence scale (\citep{IstatHBS}). 
Given that the data disclosure policies of ISTAT cannot release geo-referenced small area data, i.e. for municipality or provinces (\citep{ADELE}), we consider administrative provinces as small areas  and mimic small areas similar to provinces as follows.  For each region we set a number of clusters equal to the number of provinces and we use the amount of Waste Tax  paid by each household as a variable of clustering, since the amount of this tax is decided on the basis of municipal and provincial area specific quotes. By performing a different k-means cluster analysis for each region, we are able to preserve the heterogeneity between areas obtaining a cluster per-capita equivalised mean expenditure that goes from 297.8 to 1270.9. The samples are drawn by stratified sampling with clustered areas acting as strata, and with simple random sampling of size $n_j=N_j/20$ ($j=1,\dots,107)$ within each small area. The model used is a Log-Normal distribution defined as follows:

\begin{equation*}
\begin{cases}
 \boldsymbol{\mu}_{ij}=\exp(\beta^\mu_0+\boldsymbol{\beta}^\mu_1Age_{ij}+\beta^\mu_2Citizenship_{ij}+\boldsymbol{\gamma}^\mu_{j})\\
 \boldsymbol{\sigma}_{ij}=\exp(\beta_0^\sigma+\boldsymbol{\beta}_1^\sigma Age_{ij}+\beta_2^\sigma Citizenship_{ij}+\boldsymbol{\gamma}^\sigma_{j})\\
 \end{cases}
\end{equation*}
\noindent
where Age is the age of the respondent in 15 different classes, while Citizenship is a dummy variable that is 0 if the respondent is Italian, 1 if otherwise. The presented GAMLSS based on a Log-Normal distribution is compared with an EBLUP defined as: 
\begin{equation*}
\boldsymbol{\mu}_{ij}=\beta^\mu_0+\boldsymbol{\beta}^\mu_1Age_{ij}+\beta^\mu_2 Citizenship_{ij}+\boldsymbol{\gamma}_{j}^\mu
\end{equation*}
\noindent
 where  normality is reached due to the  Box-Cox transformation. 
    
\item [\textit{(B)}] \textit{DB income data}: we consider as pseudo-population the 2019 Italian Survey on Income and Living Conditions (IT-SILC). This survey is the main data source used to produce regular statistics on household income, poverty and social exclusion within the EU (\citep{EUSILCEU}). Our target variable is the net individual income. The total sample size of the IT-SILC is 33,254 after discarding  individuals with non-positive income, as done by \citep{graf2019generalized} for Spain. The measured targeted, in our study, is the poverty rate computed with a poverty line equal to $60\%$ of the income median. In this data-set we consider Italian regions (21) as small areas, separated from metropolitan areas (14) and the number of small areas is then equal to 35. From the population, we repeatedly select samples by stratified sampling with areas acting as strata. The sample size is $n_j=N_j/20$ ($j=1,\dots,35)$ within each small area. We use a GAMLSS with a Dagum distribution, with a model defined as follows:
\begin{equation*}
\begin{cases}
 \boldsymbol{\mu}_{ij}=\exp(\beta^\mu_0+\boldsymbol{\beta}^\mu_1 House_{ij}+\boldsymbol{\beta}^\mu_2Age_{ij}+\beta^\mu_3Citizenship_{ij}+\boldsymbol{\gamma}^\mu_{j})\\
 \boldsymbol{\sigma}_{ij}=\exp(\beta_0^\sigma+\beta_1^\sigma Adult_{ij}+\boldsymbol{\beta}_2^\sigma Age_{ij}+\beta_3^\sigma Citizenship_{ij} +\boldsymbol{\gamma}^\sigma_{j})\\
 \boldsymbol{\nu}_{ij}=\exp(\beta^\nu_0+\beta^\nu_1 Education_{ij}+ \boldsymbol{\beta}^\nu_2 Age_{ij}+ \beta^\nu_3 Citizenship_{ij}+\beta^\nu_4 Teen_{ij}+\boldsymbol{\gamma}^\nu_{j})
 \end{cases}
\end{equation*}
\noindent 
where House is a variable that assumes a value from 1 to 5, depending on the types of houses, Age is the  age divided into 4 classes, Citizenship  is a dummy variable defined as in the previous simulation, Education is a dummy variable with a value of 1 when the interviewee is still studying, while Teen and Adult are the numbers of those cohabiting, respectively, under the age of 15 and over the age of 16. The presented GAMLSS based on a Dagum distribution is compared with an EBLUP defined as: \begin{equation*}
\boldsymbol{\mu}_{ij}=\beta^\mu_0+\boldsymbol{\beta}^\mu_1House_{ij}+\boldsymbol{\beta}^\mu_2Age_{ij}+\beta^\mu_3Citizenship_{ij}+\boldsymbol{\gamma}_{j}^\mu
\end{equation*}
\end{itemize}
\noindent
where normality is reached due to the  Box-Cox transformation. In the following Section we refer to the different simulation frameworks above defined as \textit{\textit{(A) DB consumption}} data and \textit{(B)  DB income data}. 
 
\subsection{Measures of performance}
 In order to compare results, we consider indicators evaluating both  accuracy in terms of variability and bias. The following measures of performance averaged on areas are used: average relative bias (ARB), average absolute relative bias (AARB), average coefficient of variation (ACV) and average relative root mean square error (ARRMSE). Let us generalize the notation used in Section \ref{sec3} and define $H_{j}$ as the true general area parameter in the $j-th$ area and $\widehat{H}_{jt}$ its estimated value in the $t-th$ replication ($t=1,\dots,T$). We define time by time the estimator considered, whether model-based or design-based. 
  \begin{align*}
&ARB=\frac{1}{J}\sum_{j=1}^J\biggl\{ H_j^{-1}\biggl(\frac{1}{T}\sum_{t=1}^T\widehat{H}_{jt}\biggr)-1\biggr\}\times 100, \quad AARB=\frac{1}{J}\sum_{j=1}^J\biggl\{H_j^{-1}\biggl|\biggl(\frac{1}{T}\sum_{t=1}^T\widehat{H}_{jt}\biggr)-1\biggr|\biggr\}\times 100   \\
&ACV=\frac{1}{J}\sum_{j=1}^J\biggl(\frac{1}{T}\sum_{t=1}^T\frac{\sqrt{\widehat{MSE}_B(\widehat{H}_{jt})}}{\widehat{H}_{jt}}\biggr)\times 100, \quad ARRMSE=\frac{1}{J}\sum_{j=1}^J\biggl(\sqrt{\frac{1}{T}\sum_{t=1}^T\frac{ \widehat{MSE}_B(\widehat{H}_{jt})}{\widehat{H}_{jt}^2}}\biggr) \times 100
\end{align*}
where $\widehat{MSE}_B(\widehat{H}_{jt})$ is the bootstrap MSE defined in (\ref{eq:mseb}). To compare the efficiency of model-based estimators with that of the design-based one, we also use the average coefficient of variation reduction (ACVR) and the average relative efficiency (AEFF):
\begin{equation*}
 ACVR=\frac{1}{J}\sum_{j=1}^J\biggl\{\frac{1}{T}\sum_{t=1}^T\biggl(1-\frac{CV(\widehat{H}^{mb}_{jt})}{CV(\widehat{H}_{jt}^{dir})}\biggr)\biggr\}\times 100, \quad
AEFF=\frac{1}{J}\sum_{j=1}^J \biggl(\sqrt{\frac{1}{T}\sum_{t=1}^T\frac{\widehat{MSE}(\widehat{H}_{jt}^{dir})}{\widehat{MSE}_B(\widehat{H}^{mb}_{jt})}}\biggr)\times 100
    \end{equation*}where the superscript \textit{dir} denotes the direct estimator and \textit{mb} the model-based ones (i.e. GAMLSS or EBLUP). 
To conclude, in order to evaluate the performance of the proposed MSE estimator we report the Average True MSE (ATMSE), the Average Bootstrap MSE (ABMSE) and the Percentage Coverage Rate (PCR):
\begin{equation*}
ATMSE=\frac{1}{J}\sum_{j=1}^J\biggl\{\frac{1}{T}\sum_{t=1}^T\biggl(\widehat{H}_{jt}-H_{jt}\biggr)^2\biggr\} \text{,} \quad ABMSE=\frac{1}{J}\sum_{j=1}^J\biggl\{\frac{1}{T}\sum_{t=1}^T\widehat{MSE}_{B}(\widehat{H}_{jt}) \biggr\}\quad \text{and} \quad
\end{equation*}
\begin{equation*}
PCR=\frac{1}{J}\sum_{j=1}^J\biggl\{\frac{1}{T}\sum_{t=1}^T I\biggl(|\widehat{H}_{jt}-H_{jt}|\le 1.96 \sqrt{\widehat{MSE}_B(\widehat{H}_{jt})}\biggr)\biggr\} \times 100
\end{equation*}
\subsection{Discussion of simulation results}

\subsubsection{Model-based simulations}

\begin{figure}[H]
     \centering
     \begin{subfigure}[b]{0.7\textwidth}
         \centering
\includegraphics[width=\linewidth]{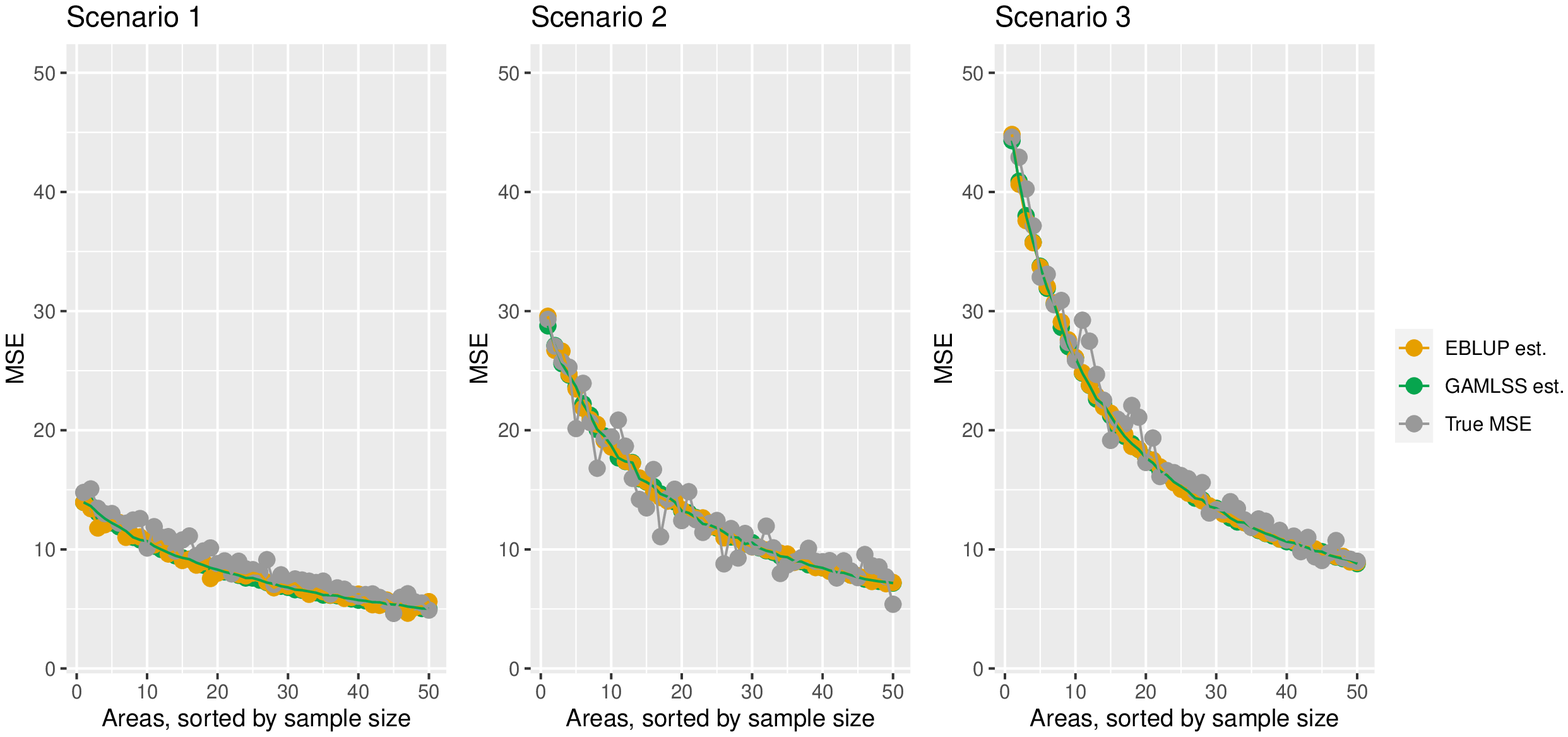}
         \caption{  \textit{MB Normal}}
         \label{fig:sima)}
     \end{subfigure}
     \hfill
          \begin{subfigure}[b]{0.7\textwidth}
         \centering
\includegraphics[width=\linewidth]{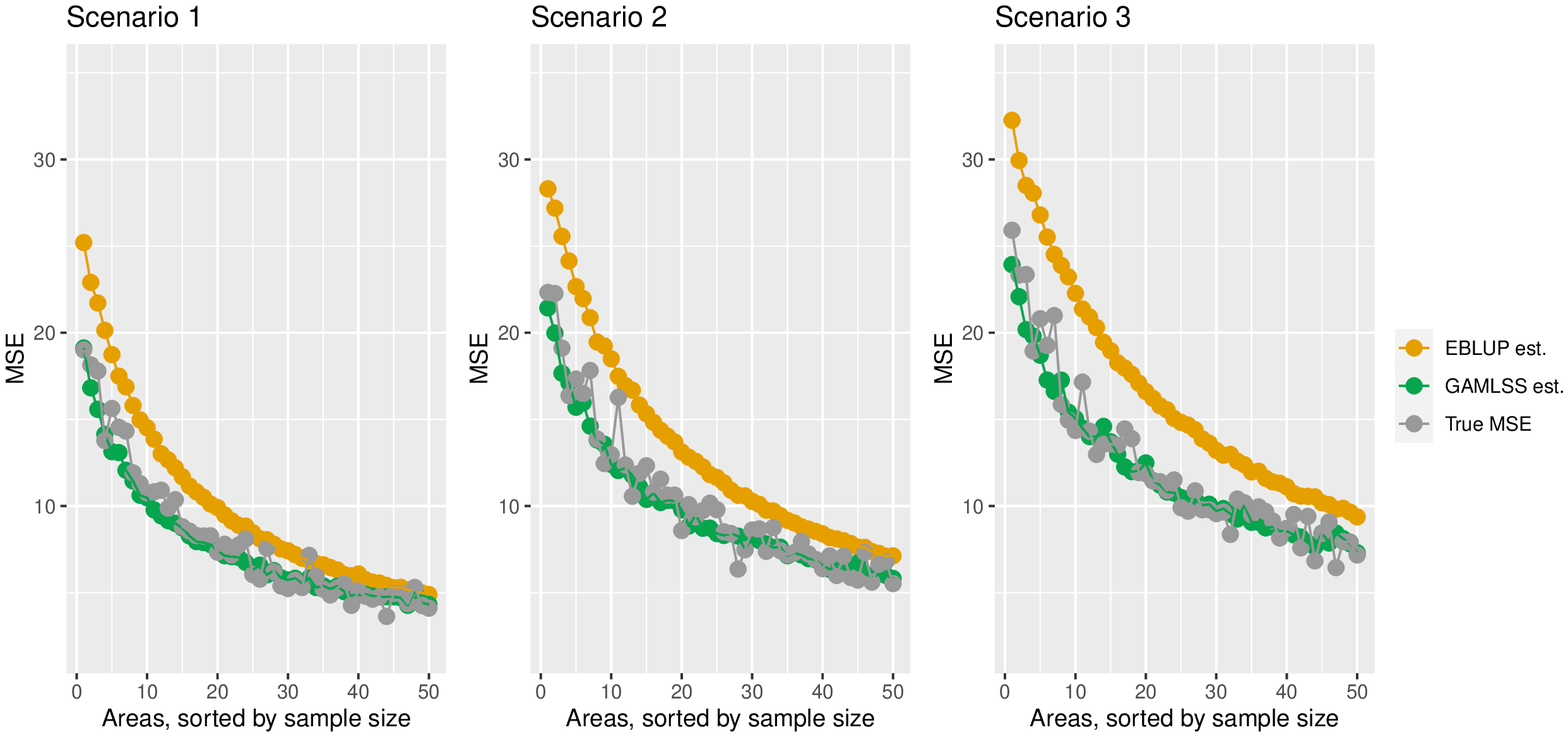}
  \caption{  \textit{\textit{MB Heteroskedastic Normal }}}
  \label{fig:simb)}
     \end{subfigure}
     \hfill
          \begin{subfigure}[b]{0.7\textwidth}
         \centering
\includegraphics[width=\linewidth]{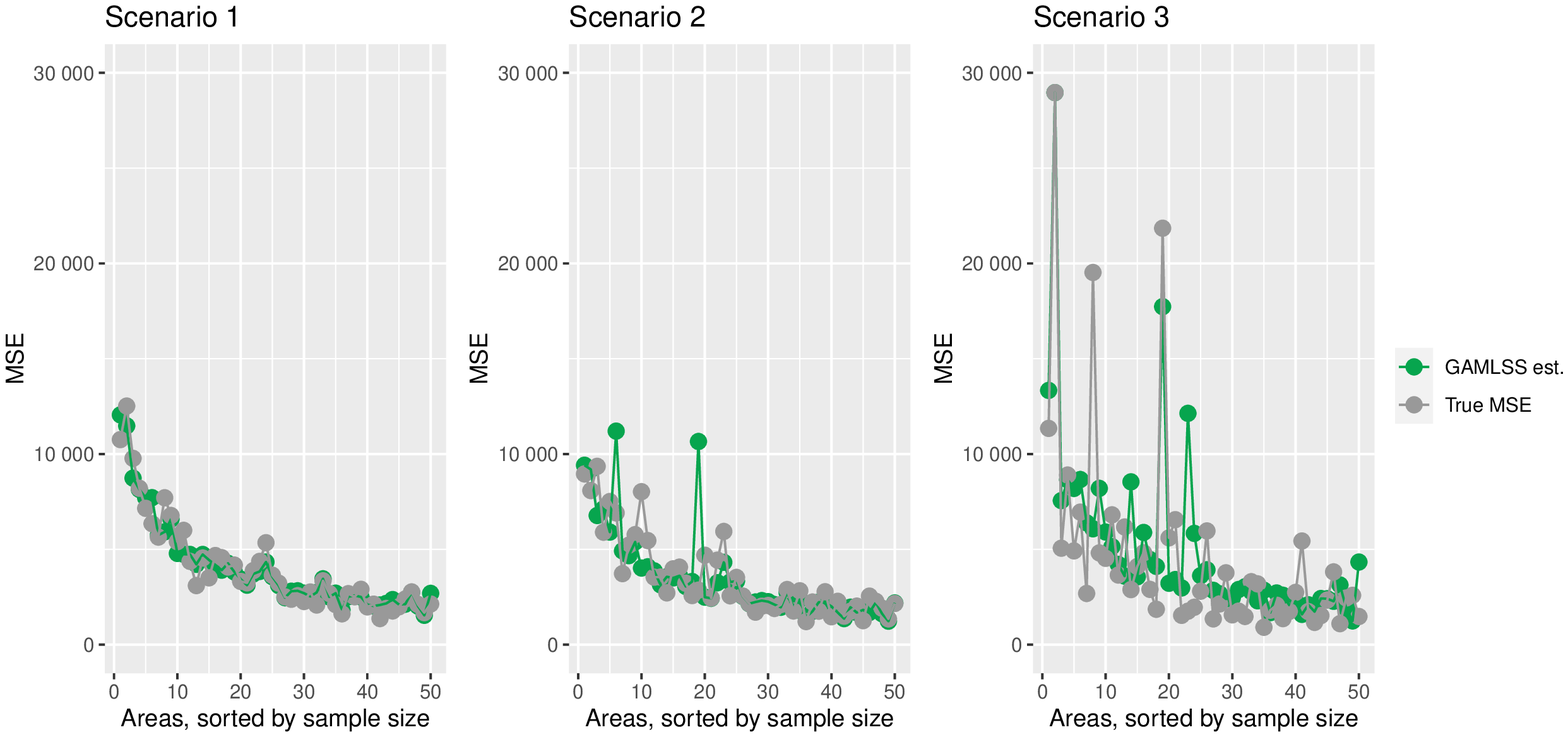}
  \caption{ \textit{MB Log-Normal}}   
  \label{fig:simc)}
     \end{subfigure}
     \hfill
          \begin{subfigure}[b]{0.7\textwidth}
         \centering
\includegraphics[width=\linewidth]{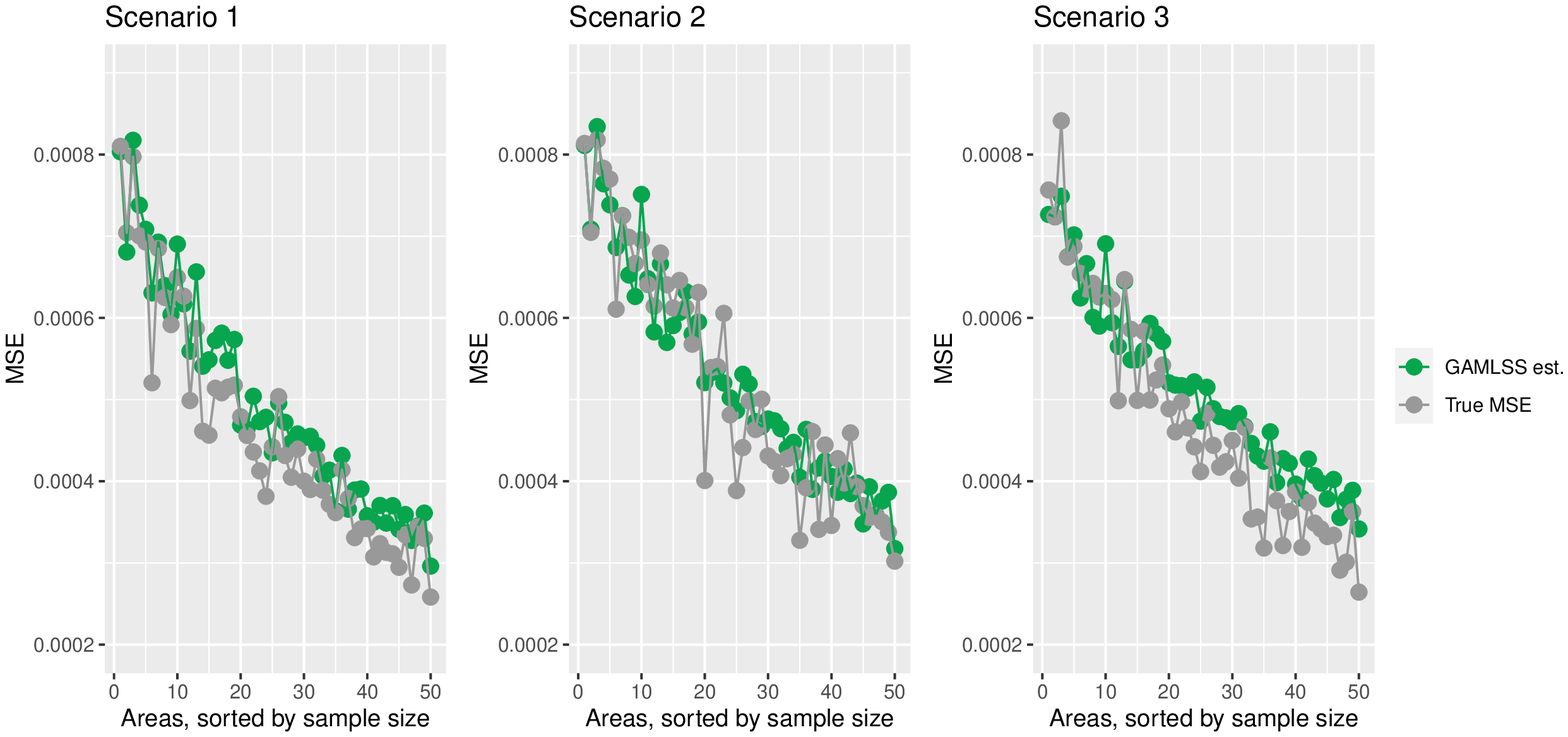}

  \caption{\textit{MB Dagum}}
  \label{fig:simd)}
     \end{subfigure}
     \hfill
\caption{
\label{multiple1}%
MSE estimation: model-based simulations}
\end{figure}

The results of measures of performance for model-based simulations are reported in Table \ref{allsim}. Figures \ref{app1} and \ref{app2}, available in Appendix A, show respectively area specific relative bias and CV. As expected, differences between GAMLSS and EBLUP are almost fully negligible in terms of bias  when data are Normal\footnote{In order to reduce the bias of $\widehat{\sigma}$ we use a correction factor as suggested by \citep{Nelder}. This is needed to take into account the error term present in $\mu$ due to the identity link function used for this parameter.} (simulations \textit{(A) MB Normal} and \textit{(B)  \textit{MB Heteroskedastic Normal}}).  Besides, results are almost equal among scenarios and some negligible differences between GAMLSS and EBLUP are given by the different method for estimating parameters and random effects. The slight increase in precision measured on average by ARB and AARB is to be sought in the outer and inner iteration of the algorithm used by GAMLSS in which $\mu$ and $\sigma$ are estimated simultaneously. Figures \ref{fig1appendix}(\subref{CompNormBias})and \ref{fig2appendix}(\subref{CompNorm}), in Appendix A,  show  that the bias and the CV for the GAMLSS and EBLUP are almost identical when data are homoskedastic. On the other hand, if data are heteroskedastic, Figures \ref{fig1appendix}(\subref{CompHetB}) and \ref{fig2appendix}(\subref{CompHet}), GAMLSS are able to reduce the CV without any change in the bias. What is important to note is the difference in the MSE estimation between simulation \textit{(A) MB Normal} and \textit{(B)  \textit{MB Heteroskedastic Normal }}. The values of the ATMSE and, ABMSE (Table \ref{allsimmse}) are very similar in both the simulations and in all the scenarios, while the values of the PCR seem to confirm the accuracy of the proposed bootstrap estimator. Area specific MSE estimation of GAMLSS and EBLUP estimators is compared with the true MSE in Figures \ref{multiple1}(\subref{fig:sima)}) and \ref{multiple1}(\subref{fig:simb)}). Results for simulation \textit{(A) MB Normal} clearly show that the MSE estimator substantially replicates the EBLUP MSE obtained by using the well-known EBLUP MSE estimator (\citep{rao_small_2015}). In simulation \textit{(B)  \textit{MB Heteroskedastic Normal }} the results on estimated MSE are completely different. Figure \ref{multiple1}(\subref{fig:simb)}) shows how the MSE GAMLSS estimators correctly estimate the MSE, significantly reducing the over-estimation that is typical of the EBLUP estimators. The reduction is more marked as the variance of the random effects increases, that is moving from the first to the third scenario. In short, the GAMLSS estimator performs better in terms of reliability compared to EBLUP and GAMLSS modelling $\sigma$ is  better able to estimate MSE. Through a specific area random effect used for $\sigma$, GAMLSS is able to reproduce the real variability in the data and  to suitably use it in the bootstrap MSE estimation.  

With regard to simulation \textit{(C)  MB Log-Normal}, form Table \ref{allsim} arises that the ARBs increase slightly if we switch from the direct estimator to model-based ones, as expected. However, GAMLSS show, in each scenario, a lower ARB than that of EBLUP. Moving on to the estimates' reliability, ACVs show that model-based estimators perform better than the direct estimator and GAMLSS performs better than EBLUP. The ARRMSE have the same trends as described for the ACV. Finally, both ACVR and AEFF clearly show how the GAMLSS are to be preferred to both the direct estimator and the EBLUP estimator. Figures \ref{fig1appendix}(\subref{CompBias}) and \ref{fig2appendix}(\subref{Comp}), in Appendix A, report area specific bias and CV. From  Figure \ref{fig1appendix}(\subref{CompBias}) we can appreciate how GAMLSS reduce the bias with respect to the EBLUP, which is affected by the back-transformation error. Table \ref{allsimmse} reports the values of ATMSE, ABMSE and PCR, which reveal the accuracy of the proposed bootstrap MSE estimator both in terms of MSE estimate and coverage rate.  Again, Figure \ref{multiple1}(\subref{fig:simc)}) compares the path of the real and estimated MSE for the GAMLSS, showing how well the latter approximates the true MSE. 

On simulation \textit{(D) MB Dagum}, we note that at first GAMLSS sharply reduces the bias with respect to the EBLUP. Secondly, differences in the ACVs increase when both parameters depend on covariates. ACVR and AEFF show how the benefits of using GAMLSS instead of EBLUP are more marked in the third scenario where both $\sigma$ and $\nu$ change between areas. Figures \ref{fig1appendix}(\subref{CompDag2}) and \ref{fig2appendix}(\subref{CompDag}) report area specific bias and CV (Appendix A). As previously seen for area specific results, we can appreciate both the reduction of relative bias if a GAMLSS model is used instead of an EBLUP and the tendency of the CV to decrease as the sample size increases. As noted by \citep{schluter_problem_2012}, the high negative values of the EBLUP estimator bias are due to the capacity of a SAE that assumes Normal distribution to estimate well the central trend, and to underestimate value on the tails of the distribution. When normality is assumed the predicted values are compressed around the mean with thin tails leading to underestimate inequality indices. This problem, however, is not evident in the MSE and in the CV, given the high capacity of the Normal distribution to tend to the mean values. To conclude, the values of ATMSE, ABMSE (Table \ref{allsimmse}) and Figure \ref{multiple1}(\subref{fig:simd)}), which compares the path of the real and estimated MSE for the GAMLSS, show the accuracy of the proposed MSE estimator even in this simulation.

\begin{table}[H]
\footnotesize
\centering
\begin{tabular}{p{1.4cm}p{0.75cm}p{1.1cm}p{1.1cm}p{0.75cm}p{1.1cm}p{1.1cm}p{0.75cm}p{1.1cm}p{1.1cm}}
\hline
 \multicolumn{10}{l}{\textit{(A)   MB Normal}}\\
 \hline
 & \multicolumn{3}{c}{Scenario 1}& \multicolumn{3}{c}{Scenario 2}& \multicolumn{3}{c}{Scenario 3}\\
\hline
 & Direct & GAMLSS & EBLUP &  Direct & GAMLSS & EBLUP &Direct & GAMLSS & EBLUP\\ 
  \hline
   ARB & 0.04 & 0.11 & 0.11 &  -0.09 & 0.01 & 0.03 & -0.02 & 0.13 & 0.14 \\ 
  AARB & 3.52 & 2.35 & 2.33& 3.84 & 2.88 & 2.85 & 4.19 & 3.41 & 3.39 \\ 
  ACV & 4.31 & 2.78 & 2.78 & 4.74 & 3.58 & 3.57 & 5.17 & 4.15 & 4.16 \\ 
  ARRMSE& 4.68 & 2.93 & 2.93 &  5.12 & 3.63 & 3.63 & 5.18 & 4.39 & 4.39\\ 
  ACVR& -- & 35.51 & 35.50& -- & 24.67 & 24.37 & -- & 19.29 & 19.34 \\
  AEFF& -- & 64.71 & 64.92 & -- & 75.64 & 75.97  & -- & 78.60 & 81.22 \\

   \hline
    \multicolumn{10}{l}{\textit{\textit{(B)   \textit{MB Heteroskedastic Normal }}}}\\
     \hline
 & \multicolumn{3}{c}{Scenario 1}& \multicolumn{3}{c}{Scenario 2}& \multicolumn{3}{c}{Scenario 3}\\
    \hline
    
 & Direct & GAMLSS & EBLUP &  Direct & GAMLSS & EBLUP &Direct & GAMLSS & EBLUP\\ 
  \hline
ARB & 0.01 & 0.08 & 0.11  &  -0.02 & 0.07 & 0.11 &  -0.06 & 0.09 & 0.11 \\ 
 AARB & 3.50 & 1.94 & 2.31 & 3.95 & 2.23 & 2.73  & 4.51 & 2.47  & 3.09 \\  
 ACV & 4.32 & 2.62 & 3.08 &  4.95 & 2.98 & 3.53 & 5.62 & 3.30 & 3.96 \\ 
ARRMSE & 4.91 & 2.84 & 3.24 &  5.87 & 3.20 & 3.73 & 6.71 & 3.50 & 4.18\\ 
 ACVR & -- & 39.08 & 28.58  & -- & 39.57 & 28.38 & -- & 41.11 & 29.29  \\ 
AEFF & -- & 61.27 & 71.87 & -- & 60.86 & 72.34 & -- & 59.63 & 71.63 \\

   \hline
    \multicolumn{10}{l}{\textit{(C)    MB Log-Normal }}\\
    \hline
 & \multicolumn{3}{c}{Scenario 1}& \multicolumn{3}{c}{Scenario 2}& \multicolumn{3}{c}{Scenario 3}\\
\hline
 & Direct & GAMLSS & EBLUP &  Direct & GAMLSS & EBLUP &Direct & GAMLSS & EBLUP\\ 
  \hline
ARB & 0.02 & -0.06 & 0.41  &   -0.04 & -0.09 & 0.26  & -0.07 & 0.01 & 0.17 \\ 
AARB & 1.96 & 1.88 & 3.64 & 1.97 & 2.07 & 3. 73 & 1.95 & 2.07 & 3.63  \\  
ACV & 21.20 & 3.81 & 6.15 &  21.31 & 3.94 & 5.05& 21.31 & 3.52 & 4.96\\ 
ARRMSE & 12.20 & 1.17 & 2.45 &  12.12& 1.31 & 2.28& 12.31 & 1.36 & 2.22  \\ 
ACVR & -- & 78.80 & 70.94 & -- & 83.05 & 76.24  & -- & 83.07 & 76.63 \\ 
AEFF & -- & 12.05  & 19.50 & -- & 14.79 &  19.47 & -- & 13.39 & 19.24 \\

    \hline
      \multicolumn{10}{l}{\textit{(D)   MB Dagum}}\\
      \hline
 & \multicolumn{3}{c}{Scenario 1}& \multicolumn{3}{c}{Scenario 2}& \multicolumn{3}{c}{Scenario 3}\\
\hline
 & Direct & GAMLSS & EBLUP &  Direct & GAMLSS & EBLUP &Direct & GAMLSS & EBLUP\\ 
  \hline
ARB & 0.13 & 0.32 & -5.34 & 0.17 & 0.33 & -6.01 & -0.05 & 0.20 & -5.83\\  
AARB & 1.80 & 4.54 & 7.23 & 1.81 & 4.86 & 7.98 & 1.81 & 4.59 & 7.34 \\ 
ACV & 23.73 & 5.34 & 6.16 & 23.75 & 5.70 & 6.42 & 23.21  & 5.05 & 7.45 \\ 
ARRMSE & 23.22 & 5.75 & 8.54 &   23.38 & 6.17 & 9.55  & 23.20 & 5.81 & 8.70 \\ 
ACVR & -- & 77.53 & 74.08  &  -- & 75.95 & 72.90& -- & 78.73 & 68.60   \\ 
AEFF & -- &  23.70 & 25.77   & -- & 25.41 & 26.79& -- & 28.00  & 26.15  \\ 
 
    \hline

\end{tabular}
\caption{Measures of performance: model-based simulations}
\label{allsim}
\end{table}

\begin{table}[H]
\footnotesize
\centering
\begin{tabular}{rrrrrrr}
\hline
 & \multicolumn{3}{l}{\textit{(A)  MB Normal}}&\multicolumn{3}{l}{\textit{\textit{(B)   \textit{MB Heteroskedastic Normal }}}}\\
 \hline
 & Scenario 1 & Scenario 2 & Scenario 3& Scenario 1 & Scenario 2 & Scenario 3\\
  \hline
 ATMSE  & 6.51  & 13.46 &  18.30 & 7.72 & 10.63 & 12.66 \\
  ABMSE  & 6.06  &  13.44  & 18.24 & 7.95   &  10.41  & 12.48   \\
  PCR   & 92.95   &  94.00   & 94.20 & 91.90   & 90.84   & 96.74 \\
  \hline
 & \multicolumn{3}{l}{\textit{(C)   MB Log-Normal} }&\multicolumn{3}{l}{\textit{(D)  MB Dagum}}\\
 \hline
 & Scenario 1 & Scenario 2 & Scenario 3& Scenario 1 & Scenario 2 & Scenario 3\\
  \hline
 ATMSE  & 4019.78 & 3510.56  & 4625.86 & 0.0004   & 0.0004 & 0.0005\\
  ABMSE  & 4023.38  &  3526.45  & 5061.31& 0.0005   &  0.0005    & 0.0005    \\
  PCR   & 91.76   &  91.57  & 92.03 & 92.03   & 91.75  & 92.84  \\
    \hline

\end{tabular}
\caption{Measures of MSE estimator performance: model-based simulations}
\label{allsimmse}
\end{table}

\noindent
\subsubsection{Design-based simulations}

\begin{figure}[H]
     \centering
     \begin{subfigure}[b]{0.8\textwidth}
         \centering
\includegraphics[width=\linewidth]{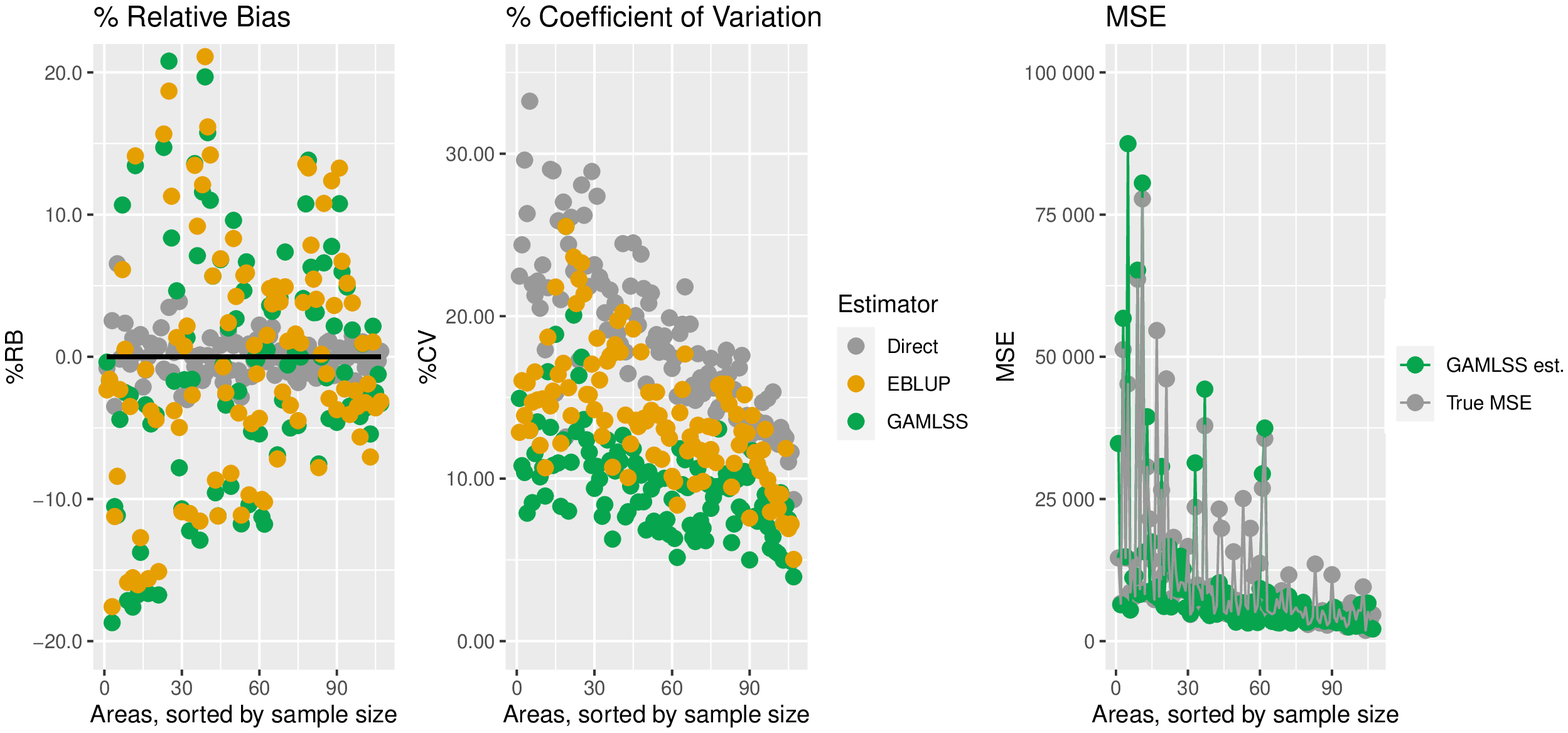}
         \caption {\textit{\textit{ DB consumption data }}}
         \label{fig:sime)}
     \end{subfigure}
     \hfill
          \begin{subfigure}[b]{0.8\textwidth}
         \centering
\includegraphics[ width=\linewidth]{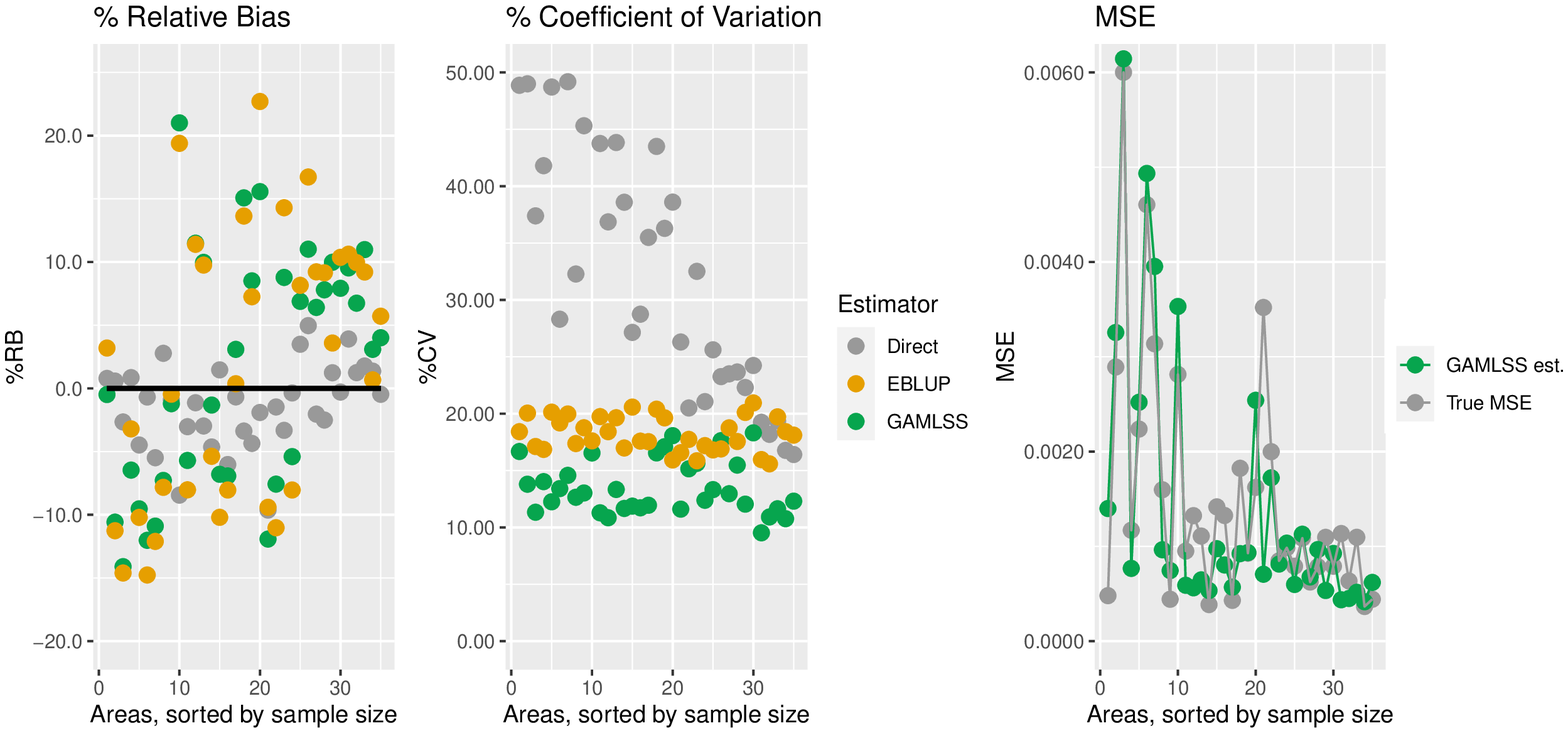}
  \caption{  \textit{DB income data }}
  \label{fig:simf)}
     \end{subfigure}
     \caption{
\label{multiple2}%
Area specific relative bias, coefficient of variation and MSE: design-based simulations}
\end{figure}

Table \ref{DisLOGNOres} shows that for each design-based simulation GAMLSS clearly reduces the AARBs if compared to the EBLUP. In simulation \textit{\textit{(A) DB consumption}} \textit{data} GAMLSS have a higher ARB than that of the EBLUP but, as highlighted by Figure \ref{multiple2}(\subref{fig:simf)}), and most of all by AARB, this is given by the compensation between positive and negative terms in the computation of the mean.  As expected, the ACV is the lowest for GAMLSS in both simulations. With regard to simulations\textit{  (A)  DB consumption data } and (B) DB \textit{income data}, ACVR shows that GAMLSS reduce the relative variability respectively by $48\%$ and 58$\%$, whereas EBLUP reduces it by only $27\%$ and 44$\%$.  From  Figure \ref{multiple2}(\subref{fig:sime)}) it also appears that EBLUP CV is for some areas even higher than those of the direct CV, and this may be due to the low $R^2$ reached with a linear mixed model. The same consideration (Figure \ref{multiple2}(\subref{fig:simf)})) can be made with regard to the CVs of simulation \textit{(B)  DB income data} where for the highest value of $n_j$ EBLUP performs worse than the direct estimator. Table \ref{allsimmseDB} reports values for the ATMSE, ABMSE and PCR. Results for the simulation \textit{\textit{(A) DB consumption}} data are satisfying and in line with those obtained for model-based simulations. On the other hand, results of \textit{(B)  DB income data} are satisfying in terms of ATMSE and ABMSE, while the not so high PCR  is due to covariates not being good (the pseudo-$R^2$ is smaller than 0.2). To conclude, from Figure \ref{multiple2} we can also appreciate that the MSE procedure defined for GAMLSS well approximates the true MSE in both simulations.

\begin{table}[H]
\footnotesize
\centering
\begin{tabular}{rrrrrrr}
\hline
&\multicolumn{3}{c}{\textit{\textit{(A) DB consumption}} \textit{data}} & \multicolumn{3}{c}{\textit{(B)  DB income data}}\\
\hline
& Direct & GAMLSS & EBLUP & Direct & GAMLSS & EBLUP \\
\hline
  ARB & -0.07 & 0.62 & 1.07 & -0.60 & 6.15 & 4.78\\ 
  AARB & 19.51 & 13.06 & 13.66 & 24.55 & 19.33 & 22.04\\ 
  ACV & 22.24  & 11.56 & 16.31 & 33.15  & 13.63  & 18.44\\ 
  ARRMSE & 26.01 & 18.42 &  19.75 & 31.61 & 22.25 & 23.94\\ 
  ACVR & -- & 47.99 & 26.57  & -- & 58.35 & 43.86\\ 
  AEFF & -- & 58.30  & 65.39 & -- & 44.99 & 60.82\\

\hline
\end{tabular}
\caption{Measures of performance: design-based simulations}
\label{DisLOGNOres}
\end{table}

\begin{table}[H]
\footnotesize
\centering
\begin{tabular}{rrr}
\hline
 & \textit{\textit{(A) DB consumption}} \textit{data} & \textit{(B)  DB income data}\\
   \hline
  ATMSE &  11964.91 & 0.0015   \\
 ABMSE &  10803.83 &  0.0008 \\
  PCR  &  91.20   &  86.74  \\
  \hline
\end{tabular}
\caption{Measures of performance of MSE estimator: design-based simulations}
\label{allsimmseDB}
\end{table}

\noindent

\section{Estimation of per-capita consumption for Italian and foreign citizens in regional rural and urban areas }
\label{sec6}
\begin{figure}[H]
\caption{Residual diagnostics plot for GAMLSS}
\label{diafnostic}
\centering
\includegraphics[width=\linewidth]{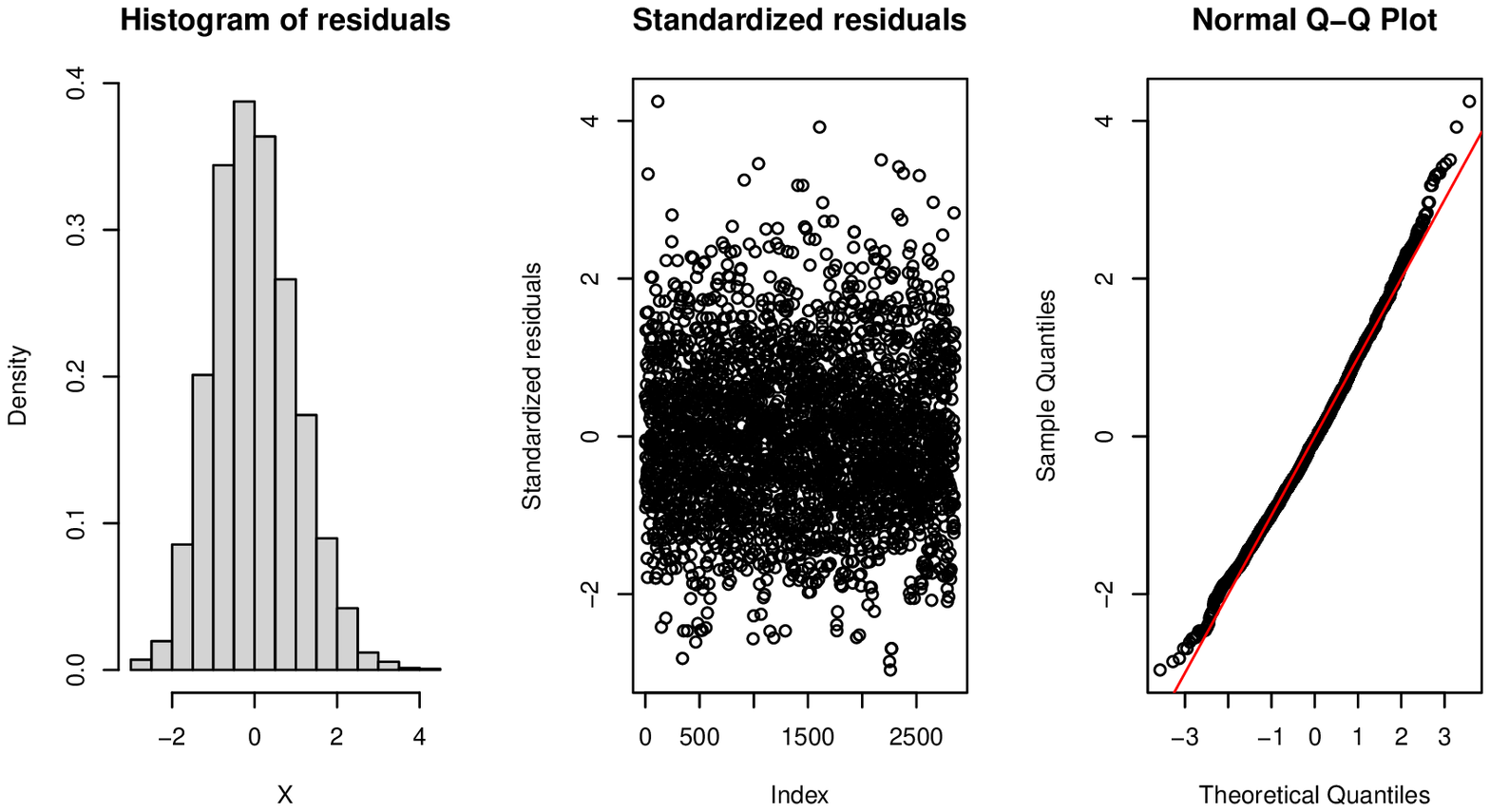}
\end{figure}

Here we use the proposed SAE model based on GAMLSS (now on SAE-GAMLSS) to estimate the per-capita equivalised consumption expenditure (PCE) of Italian and foreign citizens in the various regions of Italy, distinguishing between urban and rural areas\footnote{A person is defined as living in an urban area if lives in a community with more than 50000 citizens (\citep{dijkstra2014regional}). Two Italian regions (Valle d'Aosta and Molise) do not have any community with a populations higher than this value.}.  The reduction of the gap in the expenditure between the two groups should be a central point for every politician that aims to reduce difference and social tension between immigrants and natives.  As noted by \citep{djajic2003assimilation} and \citep{barigozzi2011immigrants}, consumption is an important dimension in the integration of immigrants in the host country. As immigrants appear to be more similar to natives, integration in terms of consumption can open up a wide range of market and social opportunities for them. 
We consider data from the 2019 Italian HBS. This survey is the main data source used to produce regular statistics on household expenditure in Italy. It gathers information on expenditure as well as on potentially auxiliary variables such as age groups and sex. We would like to point out here, that the only covariates that we are able to use are the ones known in population and available in the Adele LAB. of ISTAT. The sample sizes allow us to obtain reliable estimates of mean expenditure at a regional level. Due to small sample sizes, estimates are unreliable for foreigners with reference to regional urban and rural areas, where the minimum sample size is equal to 2 with a median number of sampled units of only 54 and direct design-based estimators for $25\%$ of the areas have a CV higher than about $21\%$. We need to restore estimates with small area estimation methods. In order to select, among a certain number of parametric distributions,  the one that best fit our data, we use the function \textit{fitDist} of the GAMLSS package (\citep{GAMLSS2019}, Ch. 6) that use the penalized maximum likelihood. The final marginal distribution is selected based on the GAIC. Following Section \ref{sec:inf}, we decide to use a GAIC with a penalty equal to  $\sqrt{\log(n)}\simeq 2$ and to also report the Bayesian information criterion (BIC). We will now move on to the following distributions: the Gamma distribution used by \citep{battese1981estimation}, the Dagum used by \citep{prieto2007spanish}, the Log-Normal used by \citep{battistin2009consumption} and, to conclude, the Student-\textit{t} used by \citep{ubaidillah2018comparative}. Alongside these distributions that have already been used by some authors to model the PCE, we also decide to include the following distributions: Normal, Skew-Normal, Skew-\textit{t}, GB2 and Singh-Maddala. Note that the GB2 distribution include as special cases the Singh-Maddala and the Dagum distributions. 
Table \ref{summaryresdist} summarises the GAIC and the BIC of the selected distributions. The two criteria  give different results: the minimum GAIC is reached with the GB2 distribution while the minimum BIC with the Log-Normal one. Based on the principle of thrift and on the properties highlighted by \citep{battistin2009consumption} for consumption distribution, we decide to use a GAMLSS based on a Log-Normal distribution with log-link function both for $\mu$ and $\sigma$. As proof of our choice, following Section \ref{sec:inf}, if the value of $k$ is increased the Log-Normal distribution has a lower GAIC than the GB2.  Figure \ref{diafnostic} and Table \ref{summaryres} show that the Log-Normal function fits our data quite well. 
\begin{table}[H]
\footnotesize
\centering
\begin{tabular}{lll}
\hline
Dist. & GAIC  & BIC\\
\hline
Dagum & 45883 & 45881  \\
Gamma & 46096 & 46108\\
GB2 & \textbf{45859} & 45882\\
Log-Normal& 45883 & \textbf{45864}\\
Normal & 47538 & 47526 \\
Singh-Maddala & 45887 & 45868 \\
Skew-Normal & 47546  & 47582 \\
Skew-\textit{t} & 45909 & 45891 \\
Student-\textit{t}& 46581 & 46563  \\
 \hline
\end{tabular}
\caption{GAIC and BIC of selected distributions}
\label{summaryresdist}
 \floatfoot{Values in bold correspond to the minimum of GAIC and BIC.}
\end{table}
\noindent

\begin{table}[H]
\footnotesize
\centering
\begin{tabular}{ll}
\hline
Mean   =  -0.001 & Coef. of skewness  =  0.297\\
 Variance   =  1.000 & Coef. of kurtosis  =  3.018\\
 \hline
\end{tabular}
\caption{Summary of the quantile residuals}
\label{summaryres}
\end{table}
\noindent
As covariates, for $\mu$, we use sex (0 for men and 1 for women) and 14 age classes, for $\sigma$, 14 age classes and, as an additive term, we consider a random effect for both parameters. To clarify, we define a SAE-GAMLSS  where the distribution model $\mathcal{F}$ is a Log-Normal model with parameters defined as follows: 
\begin{equation}
\label{modelapp}
    \begin{cases}
   \boldsymbol{\mu}_{ij}=\exp({\beta}^\mu_0+ {\beta}^\mu_1 \text{Sex}_{ij} + \boldsymbol{\beta}_1^\mu \text{Age}_{ij}+ \boldsymbol{\gamma}_{j}^\mu)\\
   \boldsymbol{\sigma}_{ij}=\exp({\beta}_0^\sigma + \boldsymbol{\beta}_1^\sigma \text{Age}_{ij}+ \boldsymbol{\gamma}_{j}^\sigma)\\
    \end{cases}
\end{equation}
\noindent
where sex and age are factors. To assess the goodness of fit of the model  (\ref{modelapp}), we monitor GAIC, BIC and the generalized log-likelihood ratio test (LR). We compare the model  (\ref{modelapp}) with a Log-Normal model without covariates, a Log-Normal model where only $\mu$ is defined by covariates and, to conclude, a Log-Normal model without random effects. 

\begin{table}[H]
\footnotesize
\centering
\begin{tabular}{lrrrrr}
\hline

\multicolumn{6}{l}{Fitted method}\\

\hdashline
 & Estimate & Std. Error & t value & Pr($>$$|$t$|$) \\ 
 
\hdashline
  
  Mu link function:  log &&&&&\\
Mu Coefficients: &&&&&\\
Intercept & 6.94 & 0.02 & 344.71 & 0.00 & $^{***}$\\ 
  Sex==W & 0.04 & 0.02 & 2.47 & 0.01 & $^{*}$\\ 
  Age 18-24 & 0.17 & 0.03 & 5.00 & 0.00 & $^{***}$ \\ 
  Age 25-29 & 0.19 & 0.03 & 5.10 & 0.00 & $^{***}$\\ 
  Age 30-34 & 0.22 & 0.04 & 6.63 & 0.00 & $^{***}$\\ 
  Age 35-39 & 0.21 & 0.03 & 6.27 & 0.00 & $^{***}$\\ 
  Age 50-44 & 0.23 & 0.03 & 6.65 & 0.00 & $^{***}$\\ 
  Age 45-49 & 0.26 & 0.04 & 7.53 & 0.00 & $^{***}$\\ 
  Age 50-54 & 0.28 & 0.03 & 8.04 & 0.00 & $^{***}$\\ 
  Age 55-59 & 0.23 & 0.04 & 5.64 & 0.00 & $^{***}$\\ 
  Age 60-64 & 0.21 & 0.06 & 3.56 & 0.00 & $^{***}$\\ 
  Age 65-69& 0.45 & 0.08 & 5.35& 0.00 & $^{***}$\\ 
  Age 70-74& 0.34 & 0.10 & 3.33 & 0.00 & $^{***}$\\ 
  Age 75+& 0.27 & 0.11 & 2.37 & 0.01 & $^{*}$\\ 
  \hdashline

    Sigma link function:  log &&&&&\\
Sigma Coefficients: &&&&&\\
  Intercept & -0.78 & 0.03 & -28.60 & 0.00 & $^{***}$\\ 
   Age 18-24 & -0.06 & 0.05 & -1.29 & 0.25 \\ 
  Age 25-29 & -0.07 & 0.06 & -0.13 & 0.90 \\ 
  Age 30-34 & 0.10 & 0.05 & 2.12 & 0.03 & $^{*}$\\ 
  Age 35-39 & 0.09 & 0.05 & 1.88 & 0.06 & $^{.}$\\ 
  Age 50-44 & 0.15 & 0.05 & 3.04 & 0.00 & $^{***}$\\ 
  Age 45-49 & 0.09 & 0.05 & 1.62 & 0.09 & $^{.}$\\ 
  Age 50-54 & 0.04 & 0.05 & 0.71 & 0.48 & \\ 
  Age 55-59 & 0.01 & 0.06 & 0.23 & 0.81 & \\ 
  Age 60-64 & 0.15 & 0.08 & 1.96 & 0.04 & $^{*}$\\ 
  Age 65-69 & 0.31 & 0.09 & 3.19 & 0.00 & $^{***}$\\ 
  Age 70-74 & -0.01 & 0.16 & -0.07 & 0.95 \\ 
  Age 75+ & 0.33 & 0.13 & 2.51 & 0.01 & $^{*}$\\ 
\hdashline
Signif. codes:&  0 $^{***}$ & 0.001 $^{**}$ & 0.01 $^{*}$ & 0.05 $^{.}$ & 0.1 $^{}$ \\
\hdashline
\multicolumn{6}{l}{No. of observations in the fit:  2854 }\\
\multicolumn{6}{l}{Degrees of Freedom for the fit:  89.84}\\
\multicolumn{6}{l}{Residual Deg. of Freedom:  2764.15 }\\
%\multicolumn{6}{l}{At cycle:  50 }\\
\hline
\end{tabular}
\caption{Model estimates}
\label{summary2}
\end{table}

\begin{table}[H]
\footnotesize
\centering
\begin{tabular}{lrrrlr}
\hline
Parameter &  Estimate &  Std. Error &   t value &  Pr($>|t|)$ & \\
 $\sigma_\mu$ & 0.22 & 0.04 & 5.28 & $<5e-7$& $^{***}$\\
 $\sigma_\sigma$ & 0.21  & 0.04 & 4.53 & $0.00$& $^{***}$\\
 \hdashline
Signif. codes:&  0 $^{***}$ & 0.001 $^{**}$ & 0.01 $^{*}$ & 0.05 $^{.}$ & 0.1 $^{}$ \\
\hline
\end{tabular}
\caption{Summary of the random effects standard deviations }
\label{summarydist}
\end{table}

\begin{figure}[H]
\caption{Estimated per-capita expenditure of foreigners and coefficient of variation }
\label{app1}
\centering
\includegraphics[height=6cm, width=\linewidth]{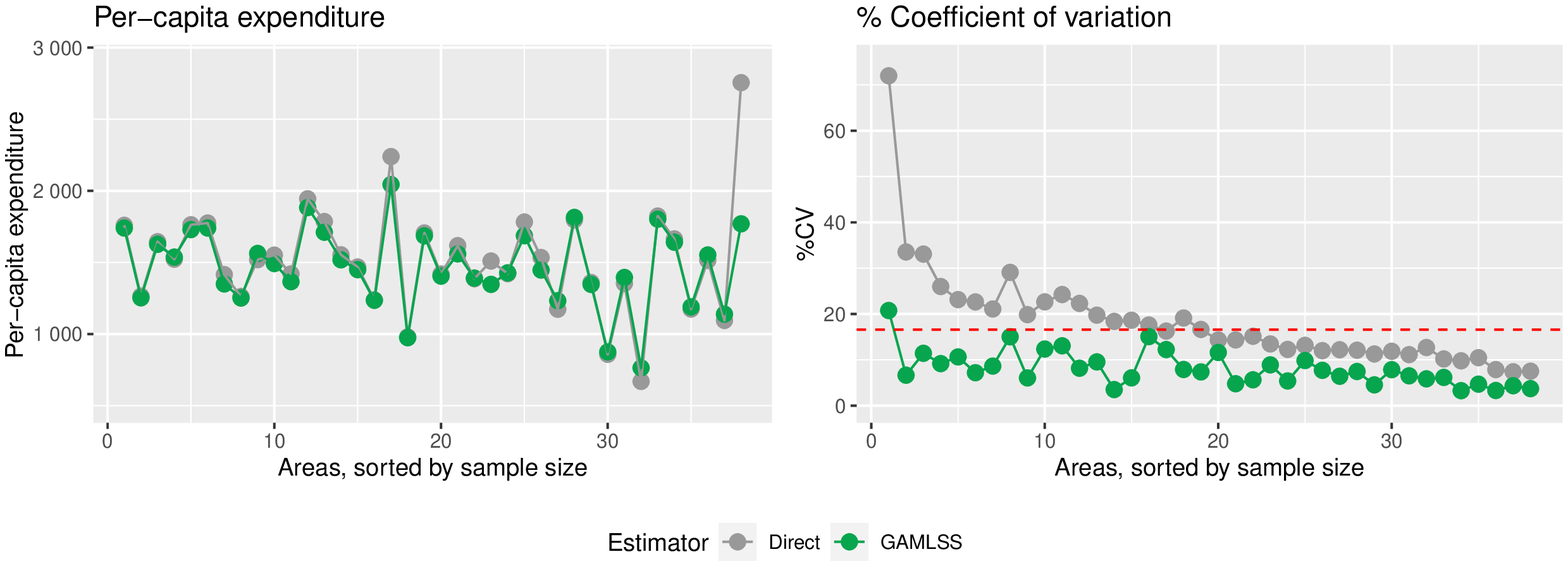}
\end{figure}

\noindent
Table \ref{summary2} reports the estimated coefficients for $\mathbf{\mu}$ and $\mathbf{\sigma}$, while Table \ref{summarydist} is about the variance of the random effects. What is important to note, with regard to the SAE frameworks, is that the possibility to define each parameter in terms of covariates, improves the model. Besides, the variances of the random effects are statistically different from 0. SAE-GAMLSS estimates of PCE for the selected domains are summarised in Figure \ref{app1}, while in Table \ref{result}  summary statistics on CV are reported for direct and SAE-GAMLSS estimators. The model-based estimated PCE has a very similar path for the unbiased design-based estimators and the SAE-GAMLSS estimator noticeably reduces the CVs. The dashed red line in Figure \ref{app1} is the threshold beyond which the estimates would be un-publishable according to the criteria adopted by, among others, \citep{Canada}. More than half the areas have a CV higher than this line, with the direct estimators reducing only to two if we use the SAE-GAMLSS based estimator. 

\begin{figure}[H]
\caption{Confidence intervals of estimated values}
\label{bp}
\centering
\includegraphics[width=\linewidth]{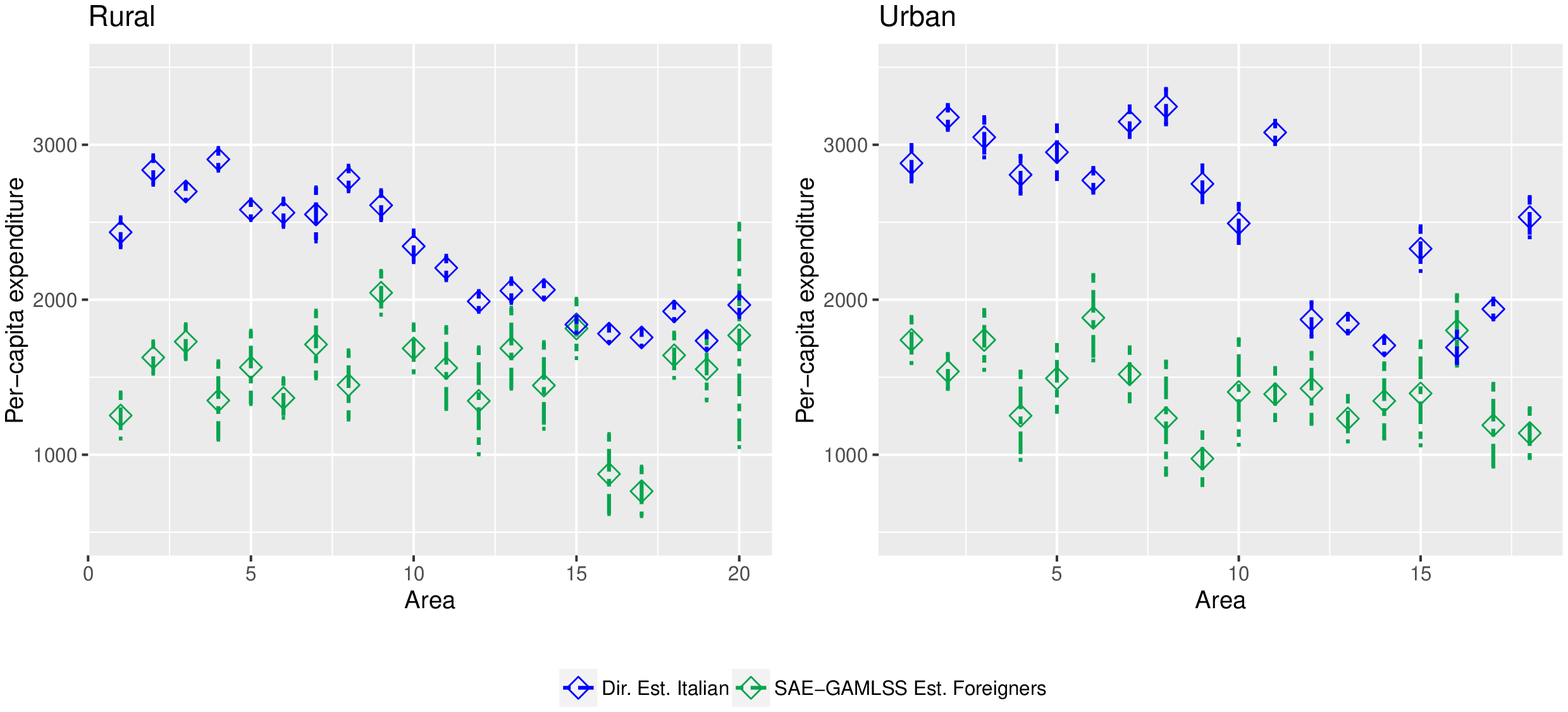}
\end{figure}

\begin{table}[H]
\centering
\begin{tabular}{rrrrrrrr}
\hline
& Min. & Q1 & Median & Mean & Q3 & Max. \\
\hline
 Direct & 7.07 & 11.48 & 15.27 & 18.18 & 20.99 & 70.77 \\ 
 GAMLSS & 3.27 & 5.70 & 7.41 & 8.13 & 9.77 & 20.76 \\
 \hline
\end{tabular}
\caption{Summary of Coefficient of Variation}
\label{result}
\end{table}
\noindent
    To obtain reliable estimates of the PCE for Italian citizens it is sufficient to use direct estimators since the mean CV is about 4.30. Figure \ref{bp} reports the $95\%$ confidence intervals of the estimated values for Italian (direct estimator) and foreign citizens (SAE-GAMLSS estimator) divided into urban and rural areas. From this figure it is possible to appreciate that in almost every  area the confidence intervals of the SAE-GAMLSS do not overlap with the ones of Italian citizens. PCE for foreigners is markedly lower than the ones of Italians. This result cannot be considered significant based on direct estimates where confidence intervals overlap.

\begin{figure}[H]
\caption{Estimated per-capita expenditure for Italian and foreign citizens}
\label{app2}
\centering
\includegraphics[width=\linewidth]{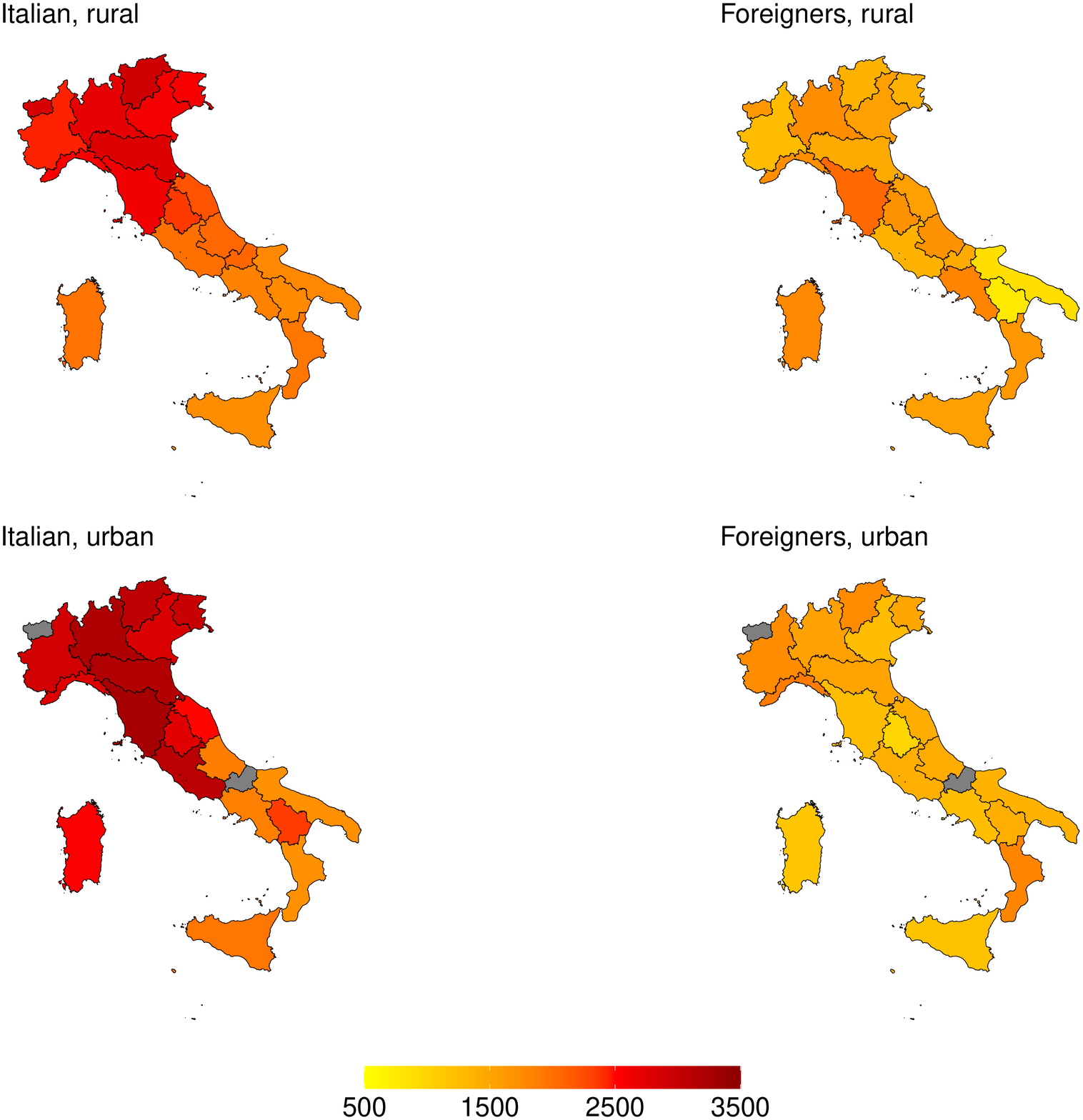}
\end{figure}

    Figure \ref{app2} maps the PCE for Italian and foreign citizens, in each regional urban and rural area. The well-known North-South divide is recognisable with reference to Italian citizens, whereas for foreigners the difference between North and South, even if present, is much less marked, both for rural and urban areas. As regards the single regions, Liguria and Toscana show the largest difference between the two groups based on citizenship, at the rural and urban level respectively. Interestingly, and perhaps unexpectedly, Apulia and Basilicata are characterised by a relative homogeneity among the four considered groups.  These considerations are only few examples based on results otherwise unobtainable without SAE models.  Reliable model-based SAE could help policy makers to define place-based policies aimed at reducing differences between Italian and foreign citizens, in this way encouraging the integration of the two groups and reducing potential social tensions.

\section{Concluding remarks}
\label{sec7}
This paper proposes a new unit-level SAE model based on GAMLSS. Under this model we present a method of obtaining small area predictions for various economic indicators on household income and consumption, and a bootstrap approach to estimate their MSE. The performances of the different estimators are compared through model and design-based simulations. Our  results provide evidences that the proposed small area predictors perform very well in terms of variability reduction with respect to the well-known EBLUP estimator. If data are normal and homoskedastic, SAE-GAMLSS and EBLUP gives almost the same results. However, in case of heteroskedasticity,  the SAE-GAMLSS  also allows us to model  $\sigma$, depending on covariates, and the related SAE predictor clearly outperforms EBLUP.  In the presence of variables with a skewed distribution, when either a Log-Normal distributional assumption is adopted or when a three-parameter Dagum distribution is assumed, the SAE-GAMLSS largely improve estimates. Finally, the strategy to estimate MSE SAE estimators is proven to be able to estimate well the true MSE. Based on the SAE-GAMLSS, we estimate the per-capita expenditure of Italian and foreign citizens, in rural and urban areas within administrative regions. The obtained maps allow us to enlighten a noticeable heterogeneity among defined domains, this allowing the definition of place-based policies aiming to  equity.

There are other areas that warrant further investigation. First of all, the implementation of the SAE-GAMLSS to estimate related outcomes could be of great interest in the economic framework (i.e. turnover and number of employees for firms, income and consumption for households). In this case the correlation between the random effects of the different outcomes can be considered in SAE-GAMLSS. Secondly, model all the parameters through covariates in GAMLSS could be helpful when the aim  in SAE is to estimate outcomes different from the mean.  An example of this is the estimation of specific non-linear indicators that, for some distributions, have a closed form. This is the case of the Gini Index which, for the Log-Normal distribution, is fully defined by the scale parameter. In this case the use of SAE-GAMLSS could allow to find the Gini index through its specific closed form. In other words, when using the Log-Normal distribution within SAE-GAMLSS, it is enough to define the scale parameter in terms of covariates and random effect to estimate the Gini index.  The main advantage is that in this way the MC approximation, usually adopted to estimate non-linear indicators in unit-level SAE models, is not necessary, this reducing the computational effort.   \\
\vspace{1cm}\\
\textbf{Acknowledgements}
We would like to thank ISTAT for kindly providing the HBS dataset made available for research purposes in the "Elementary Data Analysis Laboratory - ADELE (ADELE LAB.)".
  %  \printbibliography

\section*{Appendix A: additional simulations results}

\begin{figure}[H]
     \centering
     \begin{subfigure}[b]{0.7\textwidth}
         \centering
\centering
\includegraphics[width=\linewidth]{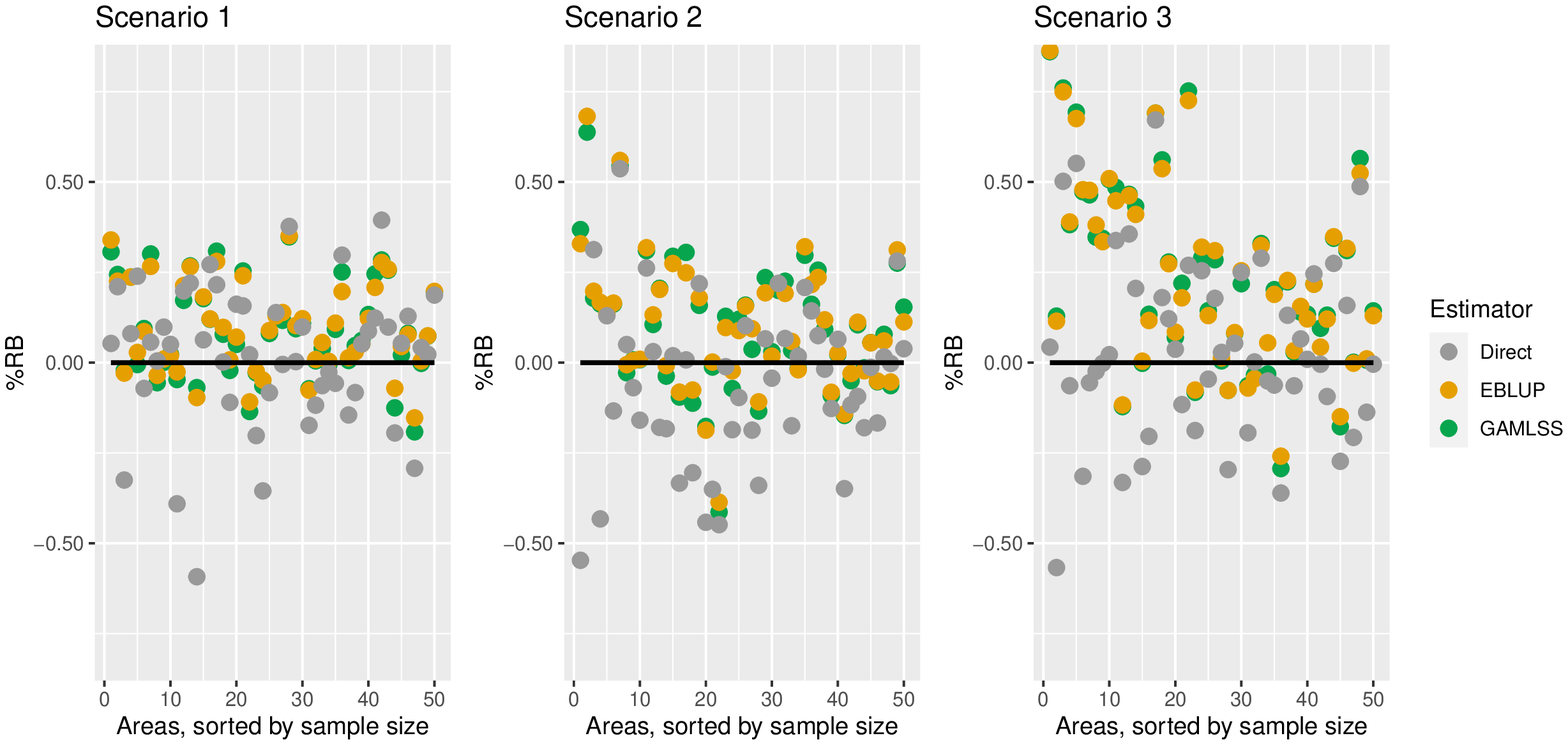}
\caption{\textit{MB Normal }}
\label{CompNormBias}
     \end{subfigure}
     \hfill
          \begin{subfigure}[b]{0.7\textwidth}
         \centering
\centering
\includegraphics[width=\linewidth]{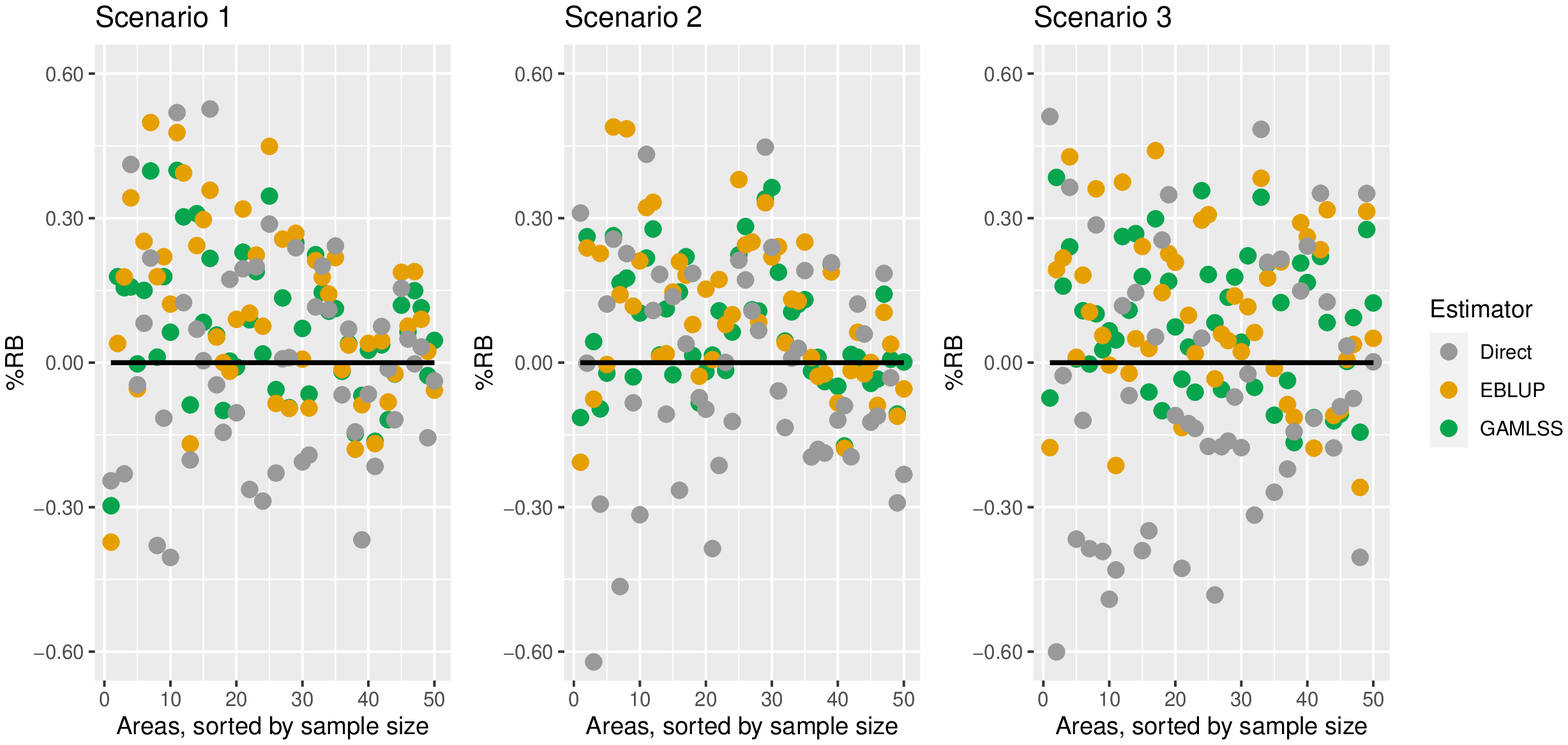}
\caption{  \textit{MB Heteroskedastic Normal } }
\label{CompHetB}
     \end{subfigure}
     \hfill
     
     \begin{subfigure}[b]{0.7\textwidth}
\centering
\includegraphics[width=\linewidth]{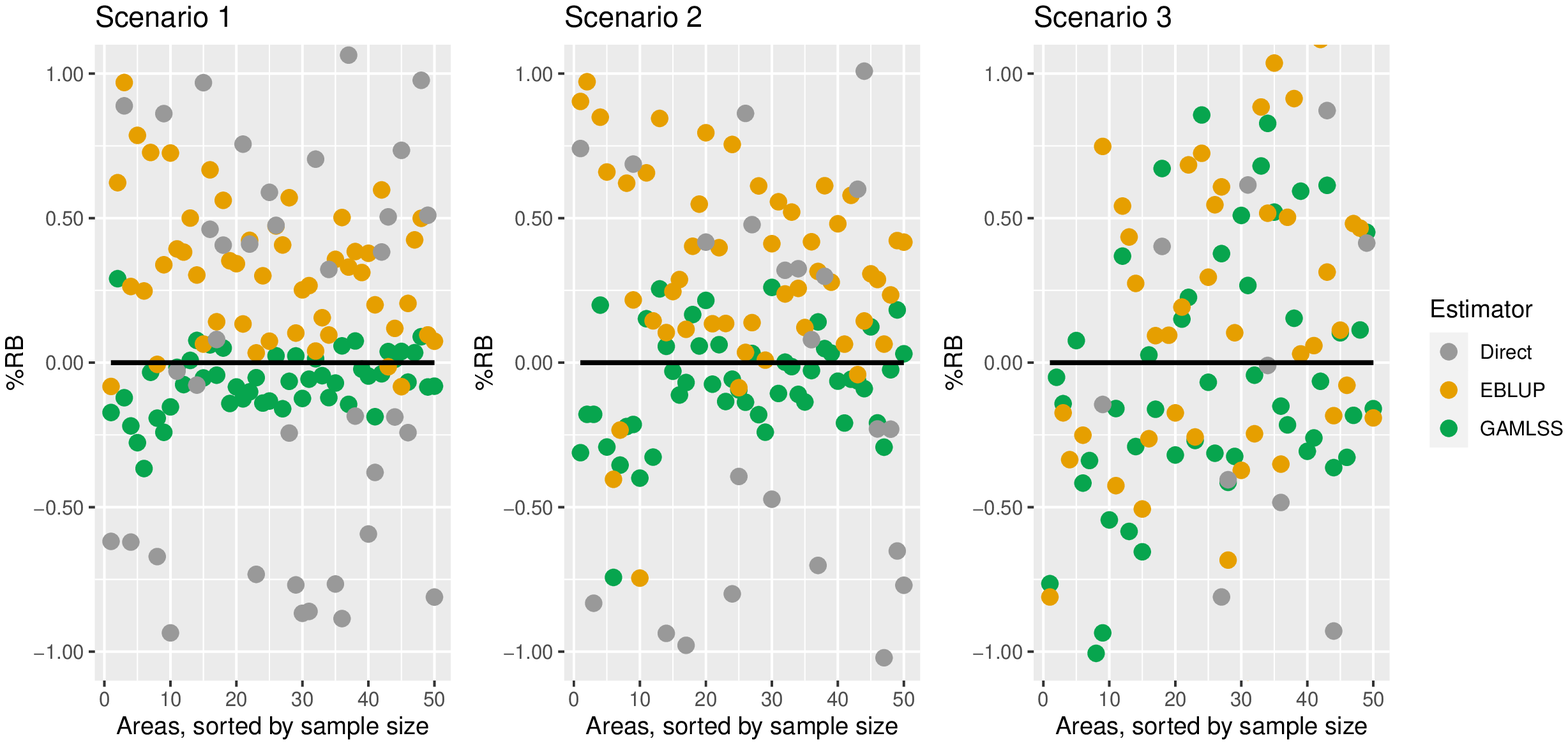}
         \caption{ \textit{ MB Log-Normal }}
\label{CompBias}
     \end{subfigure}
     \hfill
      \begin{subfigure}[b]{0.7\textwidth}
\centering
\includegraphics[width=\linewidth]{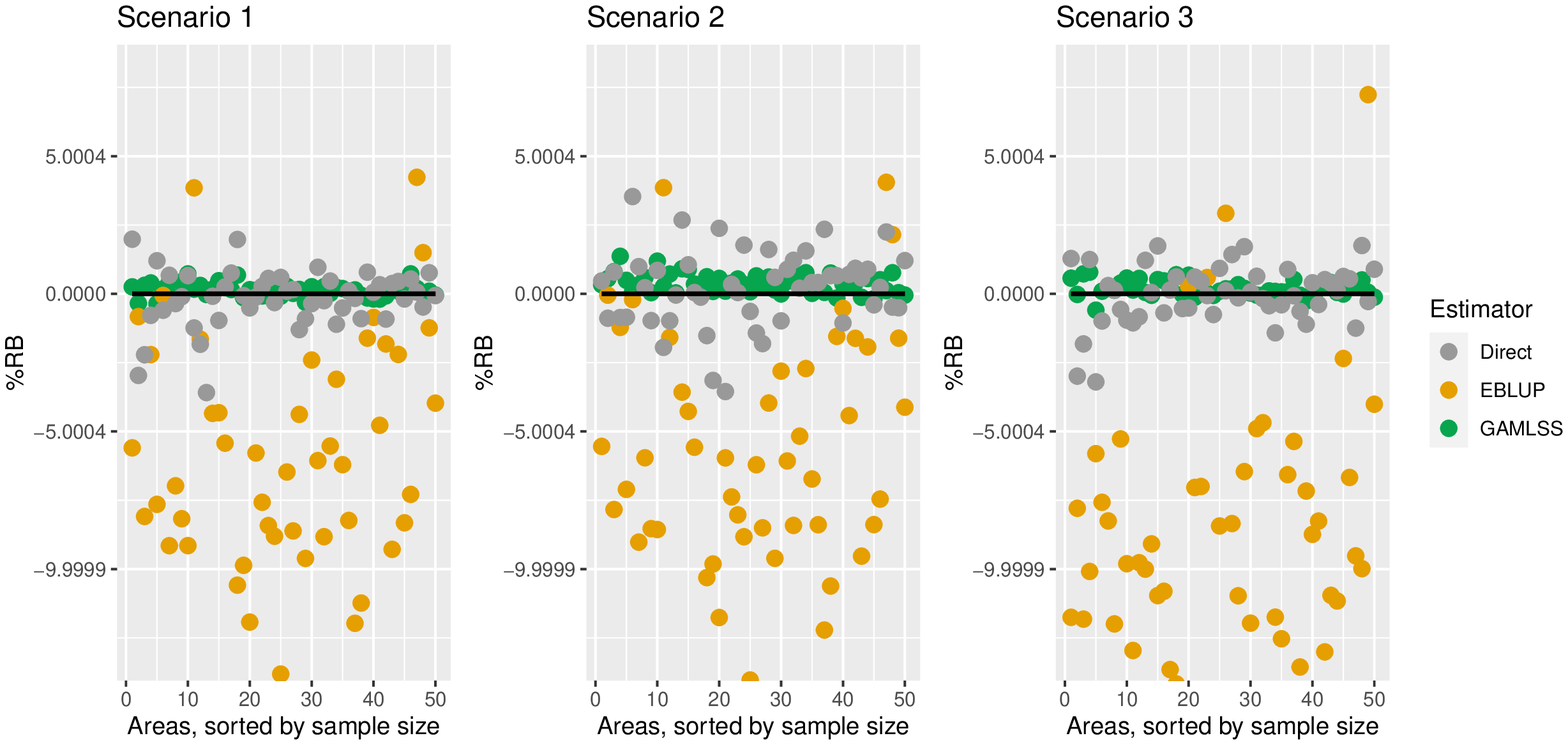}
        \caption{  \textit{  MB Dagum }}
\label{CompDag2}
     \end{subfigure}
     \hfill
     \caption{
\label{fig1appendix}%
Estimators of $\%$ relative bias: model-based simulations.}
\end{figure}

\begin{figure}[H]
     \centering
     
      \begin{subfigure}[b]{0.7\textwidth}
         \centering
\includegraphics[width=\linewidth]{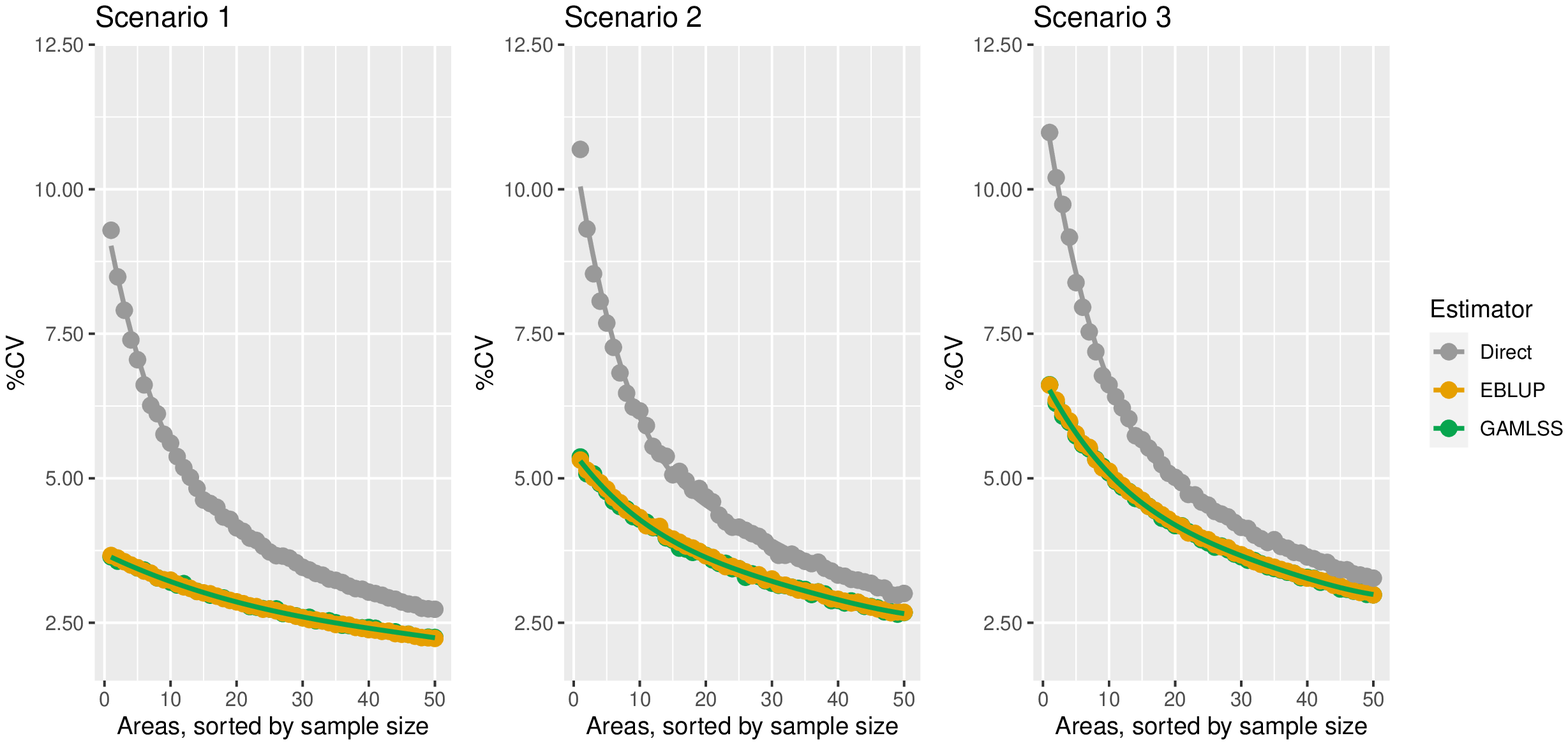}
\caption{  \textit{ MB Normal}}
\label{CompNorm}
     \end{subfigure}
     \hfill
      
     \begin{subfigure}[b]{0.7\textwidth}
         \centering

\centering
\includegraphics[width=\linewidth]{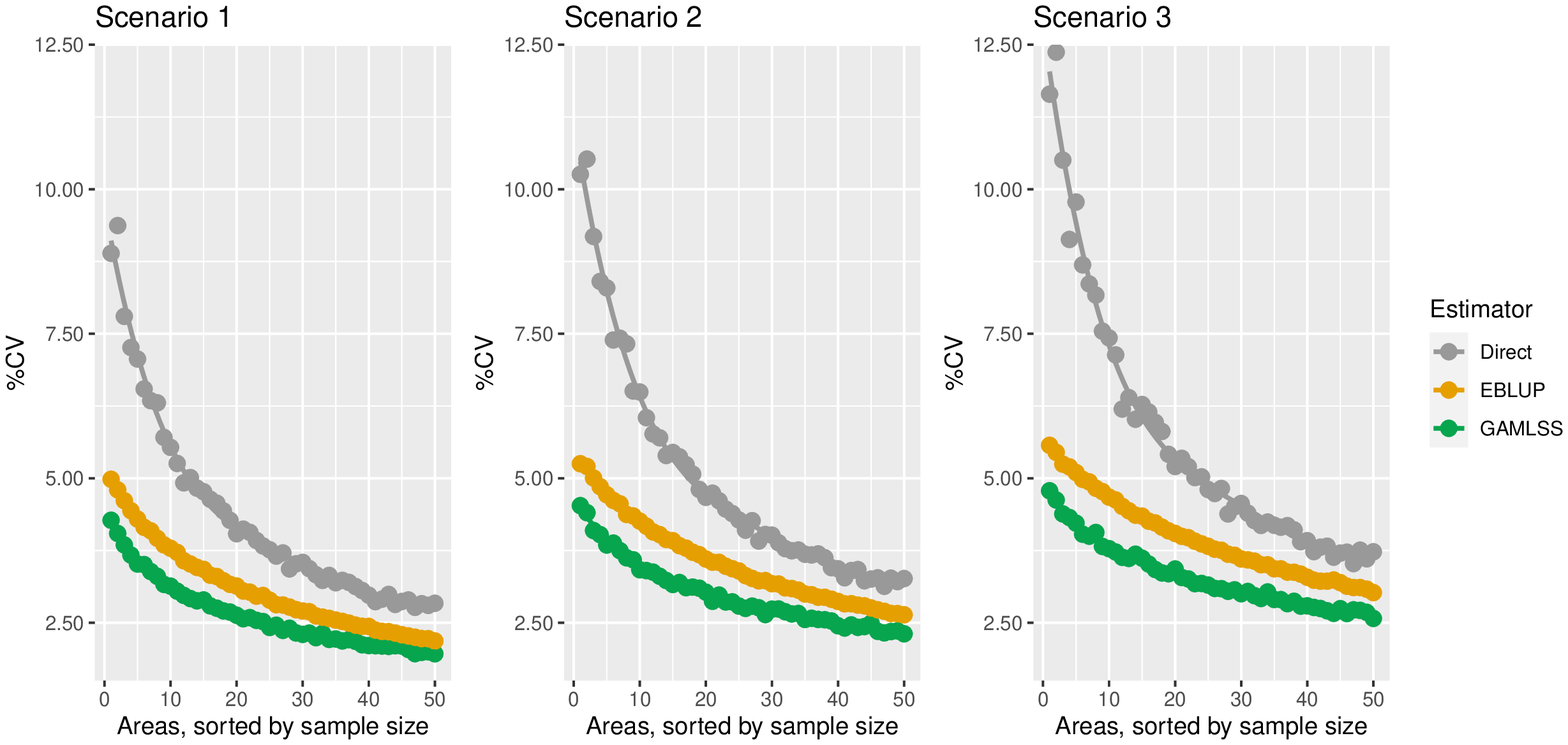}
\caption{ \textit{MB Heteroskedastic Normal }}
\label{CompHet}
     \end{subfigure}
          \begin{subfigure}[b]{0.7\textwidth}
\centering
\includegraphics[width=\linewidth]{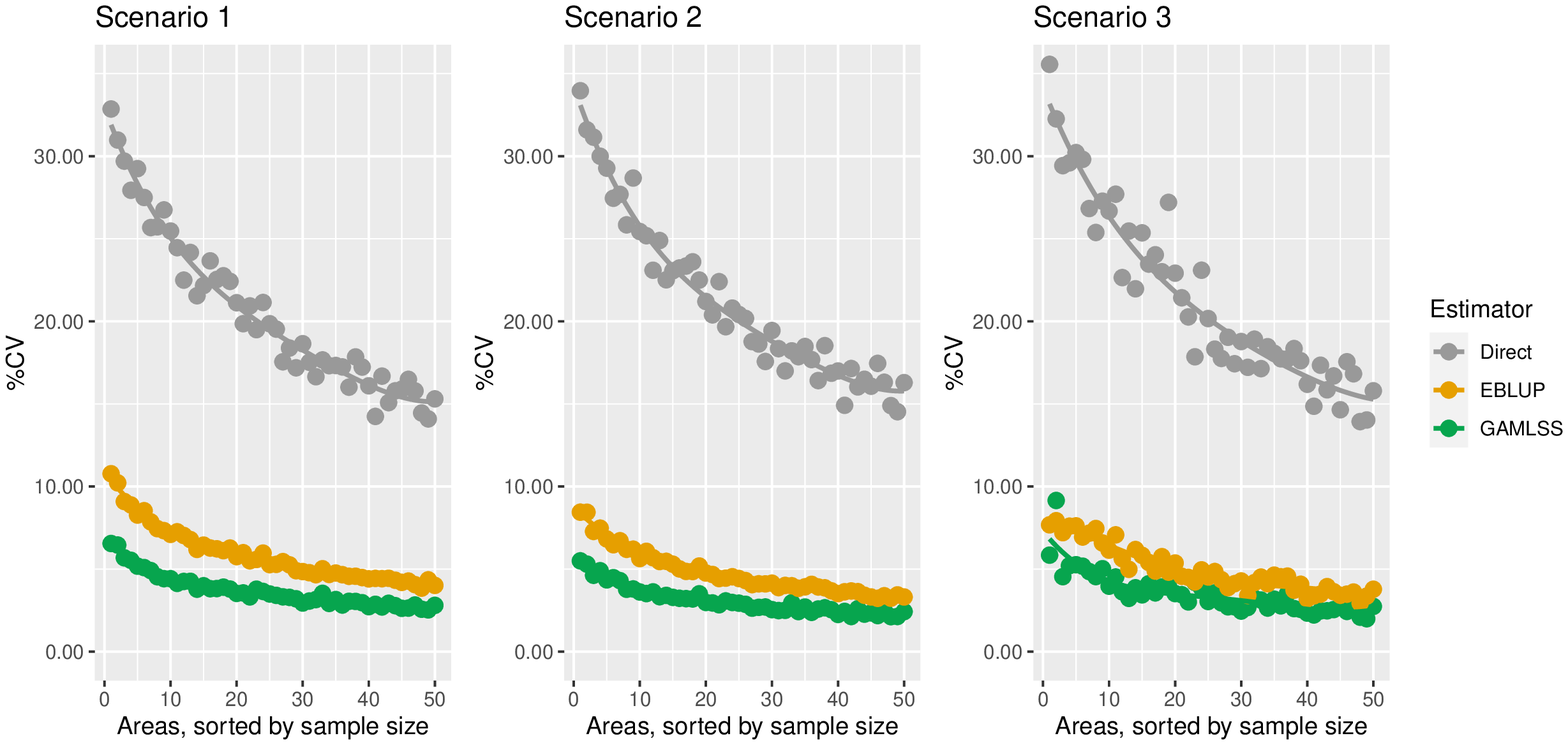}
\caption{  \textit{ MB Log-Normal }}
\label{Comp}
     \end{subfigure}
     \hfill
          \begin{subfigure}[b]{0.7\textwidth}
         \centering
\centering
\includegraphics[width=\linewidth]{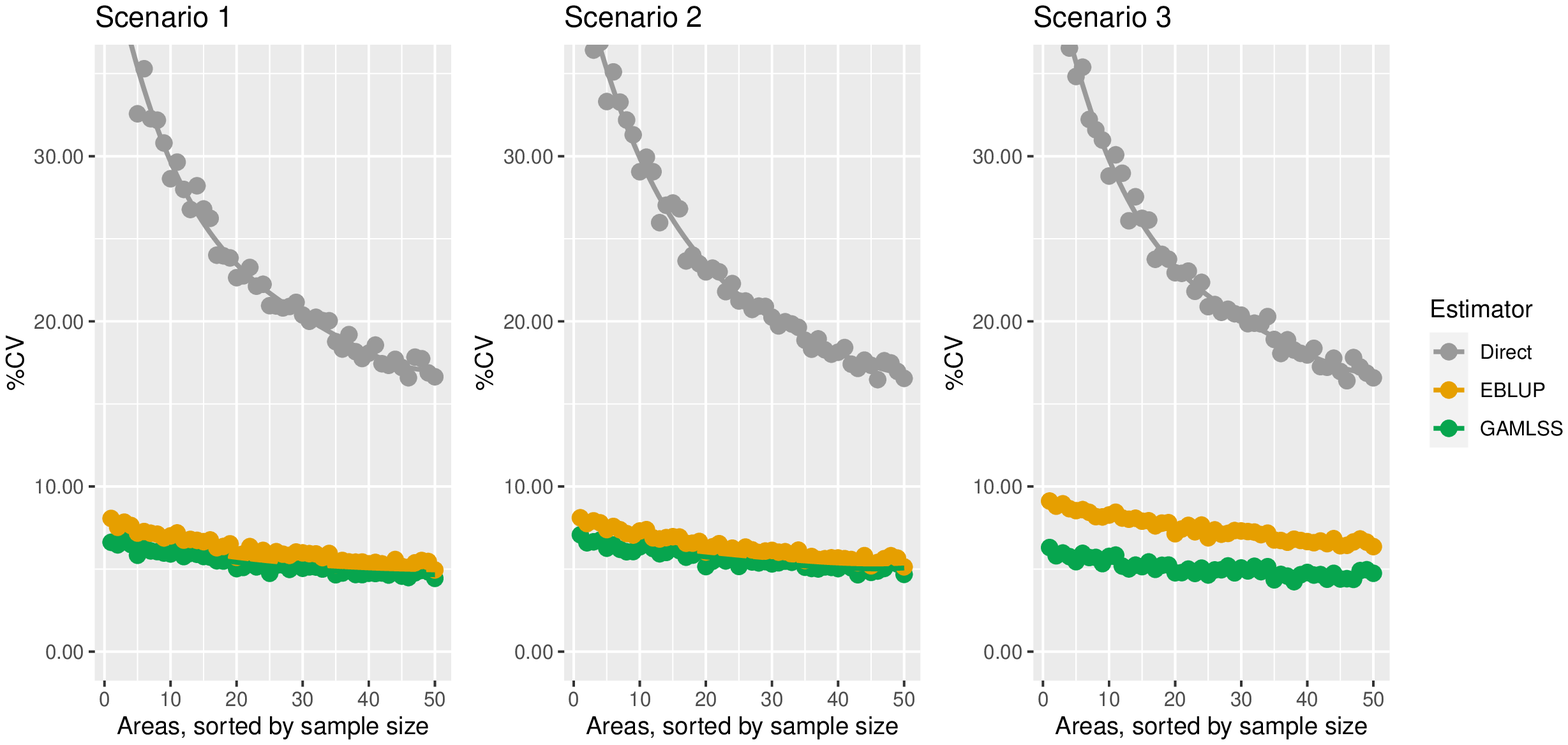}
 \caption{  \textit{ MB Dagum}}
\label{CompDag}
     \end{subfigure}
     \hfill
      \caption{
\label{fig2appendix}%
Estimator of $\%$ coefficient of variation: model-based simulations. }
\end{figure}

\end{document}